\documentclass[aps]{revtex4}
\usepackage{amssymb,epsf}

\begin{document}

\title{Dilatonic BTZ black holes with power-law field}
\author{S. H. Hendi$^{1,2}$\footnote{%
email address: hendi@shirazu.ac.ir}, B. Eslam Panah$^{1,2}$ \footnote{%
email address: behzad.eslampanah@gmail.com}, S. Panahiyan$^{1,3}$ \footnote{%
email address: sh.panahiyan@gmail.com} and A. Sheykhi$^{1,2}$\footnote{%
email address: asheykhi@shirazu.ac.ir}} \affiliation{$^1$Physics
Department and Biruni Observatory, College of Sciences, Shiraz
University, Shiraz 71454, Iran\\
$^2$ Research Institute for Astronomy and Astrophysics of Maragha (RIAAM),
Maragha, Iran \\
$^{3}$ Physics Department, Shahid Beheshti University, Tehran 19839, Iran}

\begin{abstract}
Motivated by low energy effective action of string theory and large
applications of BTZ black holes, we will consider minimal coupling between
dilaton and nonlinear electromagnetic fields in three dimensions. The main
goal is studying thermodynamical structure of black holes in this set up.
Temperature and heat capacity of these black holes are investigated and a
picture regarding their phase transitions is given. In addition, the role
and importance of studying the mass of black holes is highlighted. We will
see how different parameters modify thermodynamical quantities, hence
thermodynamical structure of these black holes. In addition, geometrical
thermodynamics is used to investigate thermodynamical properties of these
black holes. In this regard, the successful method is presented and the
nature of interaction around bound and phase transition points is studied.
\end{abstract}

\maketitle

\section{Introduction}

Appearing a scalar field in the low energy limit of string theory has
motivated various scientists to study dilaton gravity with different
viewpoints. The coupling of dilaton with other gauge fields has a profound
effects on the resulting solutions \cite%
{KoikawaY,GibbonsM,BrillH,GarfinkleHS,GregoryH,Hor2}. For example, it was
shown that the dilaton field can change the asymptotic behavior of the
spacetime. In particular, it was argued that in the presence of one or two
Liouville-type dilaton potentials, black hole spacetimes are neither
asymptotically flat nor (anti)-de Sitter ((A)dS) (see \cite%
{MignemiW,PolettiW,PolettiTW,CHM,Cai,Clem,Sheykhi0,DehghaniPH,HendiMath} and
references therein). This is due to the fact that the dilaton field does not
vanish for $r\rightarrow \infty $. Latter, it was shown that with
combination of three Liouville type dilaton potentials, it is possible to
construct dilatonic black hole solutions in the background of (A)dS
spacetime \cite{GaoZh,GaoZPLB,ShDH,HShD,Sheykhi2}. Recently, studies on
compact objects such as neutron stars in the context of dilaton gravity \cite%
{HendiBEN,Fiziev} as well as black objects in dilaton gravity's rainbow have
been carried out \cite{HendiFEP,HendiPTEP}.

On the other side, one of the important solutions of Einstein's field
equation is three dimensional black holes. The first three dimensional black
hole solution was found by Banados-Teitelboim-Zanelli (BTZ) \cite{BTZ1}, and
afterwards, BTZ black holes have got a lot of attentions \cite%
{BTZ2,BTZ3,Emparan,Hemming,Sahoo,Setare,Cadoni,Park,ParsonsR,BirminghamMS,Akbar,MyungKM,HodgkinsonL,MoonM,Frodden,EuneKY,LemosQ,Bravo,SetareA,Hosseini,WuLZ,Soroushfar}%
. The motivation to study three dimensional solutions originates from the
fact that near horizon geometry of these solutions serves as a worthwhile
model to investigate some conceptual questions of AdS/CFT correspondence
\cite{Witten07,Carlip05}. Also, studying the BTZ black hole has improved our
knowledge of gravitational systems and their interactions in 3-dimensions
\cite{Witten07}. Furthermore, possible existence of gravitational
Aharonov-Bohm effect due to the noncommutative BTZ black holes \cite%
{AnacletoB}, and specific relations between these black holes and effective
action in string theory have been explored \cite%
{Witten,LarranagaI,LarranagaII}. From the viewpoint of quantum theory of
gravity, three dimensional gravity plays an important role. Entanglement,
quantum entropy \cite{Singh,FroddenGNP,Caputa}, holography and holographical
superconductors of the BTZ black holes have been studied in Refs. \cite%
{Germani,LiuPW,de la Fuente,Chaturvedi,Ziogas,Dehyadegari}.

The studies on the BTZ black holes/wormholes were also extended to include
the dilaton field \cite{CM1,CM2}, the nonlinear electrodynamics \cite%
{BTZnon,Yamazaki,HendiJHEP,HendiEPJCthree}, and higher dimensional
spacetimes \cite{BTZlikeI,BTZlikeII}. Also, BTZ black holes in the presence
of massive gravity with Maxwell and Born-Infeld fields have been studied in
Ref. \cite{BTZmassive}. In particular, it was shown that the electric field
of BTZ-like solutions in higher-dimensions are the same as three dimensions
and they are thermally stable in the whole phase space \cite%
{BTZlikeI,BTZlikeII}. In \cite{SHS}, thermodynamical stability of static
uncharged BTZ dilaton black holes in the canonical ensemble was explored. It
was shown that depending on the dilaton coupling constant, $\alpha $, the
solution can exhibit a stable phase. Indeed, it was observed that the system
is thermally stable/unstable for specific values of the dilaton parameter.

On the other hand, gravity coupled with the nonlinear electrodynamics
attracts significant attentions because of its specific properties in
gauge/gravity coupling. Interesting physical properties of various nonlinear
electrodynamic models have been studied in many papers \cite%
{HendiJHEP,HendiEPJCthree,BTZlikeI,BTZlikeII,BTZmassive,Born,Soleng,Yajima,Delphenich,Stefanov,Gonzalez,Miskovic,Diaz-Alonso,Mazharimousavi,Kruglov,HendiEPJHEP}%
. One of the interesting branches of the nonlinear electrodynamic models is
related to power Maxwell invariant (PMI) theory in which its Lagrangian is
an arbitrary power of Maxwell Lagrangian \cite{HassaineM,Maeda,HendiE}. The
PMI theory has more interesting properties with regard to the Maxwell field,
and for unit power it reduces to linear case (Maxwell theory). One of the
attractive properties of this theory is related to the possible conformal
invariancy. It is notable that, when the power of Maxwell invariant is a
quarter of spacetime dimensions ($power = dimensions/4$), this theory enjoys
the conformal invariancy. In other words, by considering power of Maxwell
invariant equal to $dimensions/4$, one can obtain a traceless
energy-momentum tensor which leads to an invariant theory under conformal
transformation. It is notable that, considering this conformal symmetry, we
can construct charged black hole with an inverse square electric field in
arbitrary dimensions which is analogue to the four-dimensional
Reissner-Nordestr\"{o}m solutions \cite{HassaineMPRD}.

It is worth mentioning that the BTZ black holes in dilaton and dilaton
gravity's rainbow have been investigated in Refs. \cite{HendiPTEP,SHS}. Thus
it is well motivated to consider the three dimensional dilatonic black holes
in the presence of different gauge fields. Since exact solutions of charged
dilatonic black holes have not been investigate before, in the present work,
one of our goals is obtaining an exact solution of three dimensional
dilatonic black holes by adding the PMI electrodynamics to the action. In
addition, we try to study interesting properties of such exact solution as a
thermodynamical system.

One of the interesting methods for studying properties of thermodynamical
systems is through geometrical thermodynamics. This method employs
Riemannian geometry to construct phase space. The information regarding
thermodynamical behavior of a system could be extracted from the Ricci
scalar of this phase space. Divergencies of the Ricci scalar determine two
sets of important points: bound points which separate situations with
positive temperature from those with negative temperature, and phase
transition points which represent discontinuities in thermodynamical
quantities such as the heat capacity. In addition, considering the sign of
Ricci scalar around its divergencies, it is possible to determine whether
the interaction around divergencies are of repulsive or attractive nature
\cite{Rupp}.

The first geometrical thermodynamical approach was introduced by Weinhold
which is based on internal energy as thermodynamical potential \cite%
{WeinholdI,WeinholdII}. Latter, an alternative was proposed by Ruppeiner
which has entropy as its thermodynamical potential \cite%
{RuppeinerI,RuppeinerII}. Another proposal for the geometrical
thermodynamics was given by Quevedo which has Legendre invariancy as its
core stone \cite{QuevedoI,QuevedoII}. Mentioned methods have been used in
studying thermodynamics of the black holes \cite%
{HanC,BravettiMMA,Ma,GarciaMC,ZhangCY,MoLW2016,Sanchez,Soroushfar1,HendiNaderi}%
. In a series of papers, it was shown that the mentioned methods may
confront specific problems in describing thermodynamical properties of the
black holes \cite{HPEMI,HPEMII,HPEMIII,HPEMIV}. In other words, there were
cases in which, the results of these three approaches were not consistent
with those extracted from other methods. In order to remove the shortcomings
of other methods regarding geometrical thermodynamics, a new thermodynamical
metric was introduced by Hendi, Panahiyan, Eslam Panah and Momennia (HPEM)
\cite{HPEMI}. It was shown that employing this method leads to consistent
results regarding thermodynamical properties of the black holes. For a
comparative study regarding these four thermodynamical metrics, we refer the
readers to Ref. \cite{Wen}.

The organization of our paper is as follows. In the next section, we present
the field equations of three dimensional charged dilatonic black holes when
the gauge field is in the form of power Maxwell field. In section \ref{Therm}%
, we calculate conserved and thermodynamic quantities of obtained solutions
and examine the validity of the first law of thermodynamics. Thermodynamical
behavior of the solutions and the properties governing it are explored in
different contexts in sections \ref{TB} and \ref{GT}. The paper is concluded
in section \ref{Con} with some closing remarks.

\section{Field equation and Charged dilaton black hole solutions}

\label{Field}

In this section, we obtain three dimensional charged dilatonic black hole
solutions in the presence of PMI field. For this purpose, we consider a $3-$%
dimensional action of Einstein gravity which is coupled with dilaton and PMI
fields
\begin{equation}
\mathcal{I}=\frac{1}{16\pi }\int d^{3}x\sqrt{-g}\left[ \mathcal{R}-4\left(
\nabla \Phi \right) ^{2}-V\left( \Phi \right) +\left( -e^{-4\alpha \Phi }%
\mathcal{F}\right) ^{s}\right] ,  \label{action}
\end{equation}%
where $\mathcal{R}$ is the Ricci scalar, $\Phi $ is the dilaton field and $%
V\left( \Phi \right) $ is a dilatonic potential. Also, $\mathcal{F}=F_{\mu
\nu }F^{\mu \nu }$ is the Maxwell invariant, in which $F_{\mu \nu }=\partial
_{\mu }A_{\nu }-\partial _{\nu }A_{\mu }$ is the Faraday tensor with the
electromagnetic potential $A_{\mu }$ and $s$ the power of nonlinearity. In
addition, it should be pointed out that $\alpha $ is a constant which
determines the strength of coupling of the scalar and electromagnetic field.

Using variational principle and varying Eq. (\ref{action}) with respect to
the gravitational, dilaton and gauge fields ($g_{\mu \nu }$, $\Phi $ and $%
A_{\mu }$), we find the following field equations
\begin{equation}
R_{\mu \nu }=4\left( \partial _{\mu }\Phi \partial _{\nu }\Phi +\frac{1}{4}%
g_{\mu \nu }V(\Phi )\right) +e^{-4\alpha s\Phi }\left( -\mathcal{F}\right)
^{s}\left[ \left( 2s-1\right) g_{\mu \nu }-\frac{2s}{\mathcal{F}}F_{\mu
\lambda }F_{\nu }^{~\ \lambda }\right] ,  \label{dilaton equation(I)}
\end{equation}%
\begin{equation}
\nabla ^{2}\Phi =\frac{1}{8}\frac{\partial V}{\partial \Phi }+\frac{s\alpha
}{2}e^{-4s\alpha \Phi }\left( -\mathcal{F}\right) ^{s},
\label{dilaton equation(II)}
\end{equation}%
\begin{equation}
\partial _{\mu }\left( \sqrt{-g}e^{-4\alpha s\Phi }\left( -\mathcal{F}%
\right) ^{s-1}F^{\mu \nu }\right) =0.  \label{Maxwell equation}
\end{equation}

In this paper, we are attempting to obtain charged dilatonic black hole
solutions in $(2+1)$-dimensions. To do so, one can employ following static
metric ansatz
\begin{equation}
ds^{2}=-f(r)dt^{2}+\frac{dr^{2}}{f(r)}+r^{2}R^{2}(r)d\varphi ^{2},
\label{metric}
\end{equation}%
where $f(r)$ and $R(r)$ are radial dependent functions which should be
determined.

Since we are looking for the black holes with a radial electric field ($%
F_{tr}(r)=-F_{rt}(r)\neq 0$), the electromagnetic potential will be in the
following form
\begin{equation}
A_{\mu }=\delta _{\mu }^{0}h\left( r\right) .  \label{electric po}
\end{equation}

Using Eqs. (\ref{Maxwell equation}) and (\ref{electric po}), and considering
the metric (\ref{metric}), one can obtain electromagnetic tensor as
\begin{equation}
F_{tr}=E(r)=qe^{\frac{4\alpha s\Phi }{2s-1}}\left[ rR(r)\right] ^{\frac{1}{%
1-2s}},  \label{Ftr eq}
\end{equation}%
where $q$ is an integration constant which is related to the electric charge
of black holes.

Now, we can obtain field equations by using Eqs. (\ref{dilaton equation(I)})
and (\ref{dilaton equation(II)}) as
\begin{eqnarray}
eq_{tt} &:&2^{s}\left( s-1\right) \left( E^{2}(r)e^{-4\alpha \Phi
(r)}\right) ^{s}+\frac{1}{2}\left[ f^{\prime \prime }(r)+\left( \frac{1}{r}+%
\frac{R^{\prime }(r)}{R(r)}\right) f^{\prime }(r)\right] +V(\Phi )=0,\
\label{tt} \\
&&  \nonumber \\
eq_{rr} &:&eq_{tt}+\left[ \frac{R^{\prime \prime }(r)}{R(r)}+\frac{%
2R^{\prime }(r)}{rR(r)}+4\Phi ^{\prime 2}(r)\right] f(r)=0,  \label{rr} \\
&&  \nonumber \\
eq_{\theta \theta } &:&2^{s}\left( 1-2s\right) \left( E^{2}(r)e^{-4\alpha
\Phi (r)}\right) ^{s}+\left[ \frac{R^{\prime \prime }(r)}{R(r)}+\frac{%
2R^{\prime }(r)}{rR(r)}\right] f(r)+\left[ \frac{1}{r}+\frac{R^{\prime }(r)}{%
R(r)}\right] f^{\prime }(r)+V(\Phi )=0,  \label{com}
\end{eqnarray}%
where the prime and double prime are the first and the second derivatives
with respect to $r$, respectively. Subtracting the equations (\ref{tt}) of (%
\ref{rr}) ($eq_{tt}-eq_{rr}$), we can obtain the following equation
\begin{equation}
\frac{R^{\prime \prime }(r)}{R(r)}+\frac{2R^{\prime }(r)}{rR(r)}+4\Phi
^{\prime 2}(r)=0.
\end{equation}

Next, we employ an ansatz, $R(r)=e^{2\alpha \Phi (r)}$, in the above field
equation. The motivation for considering such an ansatz is due to black
string solutions of Einstein-Maxwell-dilaton gravity which was first
introduced in Ref. \cite{Dehghani}. The $\Phi (r)$\ is obtained
\begin{equation}
\Phi (r)=\frac{\gamma }{2}\ln \left( \frac{b}{r}\right) ,  \label{Phi}
\end{equation}%
where $b$\ is an arbitrary constant and $\gamma =\alpha ^{2}/K_{1,1}$\ ($%
K_{i,j}=i+j\alpha ^{2}$).

Here, in order to find consistent metric functions, we use a modified
version of Liouville-type dilaton potential
\begin{equation}
V(\Phi )=2\lambda e^{4\beta \Phi }+2\Lambda e^{4\alpha \Phi },
\label{V(Phi)}
\end{equation}%
where $\Lambda $\ is a free parameter which plays the role of cosmological
constant, $\lambda $\ and $\beta $\ are arbitrary constants. This potential
is the usual Liouville-type dilaton potential that is used in the context of
Friedman-Robertson-Walker (FRW) scalar field cosmologies \cite{Ozer} and
Einstein-Maxwell-dilaton black holes \cite{Chan,Yazadjiev,Sheykhi}.
Substituting Eqs. (\ref{Phi}) and (\ref{V(Phi)}) in (\ref{Ftr eq}) and using
the ansatz, $R(r)=e^{2\alpha \Phi (r)}$\ in the field equation (\ref{com}),
we can construct exact charged black hole solutions of this theory by
finding the metric functions, $f(r)$,\ as
\begin{eqnarray}
f(r) &=&\frac{2\mathcal{K}_{1,1}^{2}\Lambda r^{2}}{\mathcal{K}_{-2,1}}\left(
\frac{b}{r}\right) ^{2\gamma }-mr^{\gamma }  \nonumber \\
&&+\frac{\left( 2q^{2}\right) ^{s}\left( 2s-1\right) ^{2}\mathcal{K}%
_{1,1}^{2}}{\mathcal{K}_{2,1}-2s}r^{\frac{2\left( s-1\right) }{2s-1}}\left[
1+\frac{\alpha ^{2}\left( s-1\right) }{\left( \alpha ^{2}-s\mathcal{K}%
_{-1,1}\right) }\right] ,  \label{f(r)}
\end{eqnarray}%
where $m$\ is an integration constant which is related to the total mass of
black holes. It is notable that, the above solutions (Eqs. (\ref{Phi}) and (%
\ref{f(r)})) will fully satisfy the system of equations provided,
\begin{eqnarray}
\lambda  &=&\frac{\alpha ^{2}\left( s-1\right) \left( 2s-1\right)
2^{s-1}q^{2s}}{\left( \alpha ^{2}+s\mathcal{K}_{1,-1}\right) b^{\frac{2s}{%
2s-1}}}, \\
&&  \nonumber \\
\beta  &=&\frac{s\alpha }{\gamma \left( 2s-1\right) }.
\end{eqnarray}

On the other hand, one can easily show that the vector potential $A_{\mu }$,
corresponding to the electromagnetic tensor (\ref{Ftr eq}), can be written as%
\begin{equation}
A_{\mu }=\frac{\left( 2s-1\right) qb^{\gamma }\mathcal{K}_{1,1}}{2s-\mathcal{%
K}_{2,1}}r^{\frac{2s-\mathcal{K}_{2,1}}{\left( 2s-1\right) \mathcal{K}_{1,1}}%
},
\end{equation}

The electromagnetic gauge potential should be finite at infinity, therefore,
one should impose following restriction to have this property
\begin{equation}
\frac{2s-\mathcal{K}_{2,1}}{\left( 2s-1\right) \mathcal{K}_{1,1}}<0.
\end{equation}

The above equation leads to the following restriction on the range of $s$%
\begin{equation}
\frac{1}{2}<s<\frac{2+\alpha ^{2}}{2}.
\end{equation}

It is notable that, in the absence of a non-trivial dilaton ($\alpha =\gamma
=0$), the solutions (\ref{f(r)}) reduce to
\begin{equation}
\Psi (r)=-m-\Lambda r^{2}+\frac{\left( 2q^{2}\right) ^{s}\left( 2s-1\right)
^{2}}{2\left( 1-s\right) }r^{\frac{2\left( s-1\right) }{2s-1}}.  \label{PMI3}
\end{equation}%
which describes a $3$-dimensional charged black hole in Einstein-PMI
gravity, provided the PMI parameter must be in the range $\frac{1}{2}<s<1$.
It is notable that, for $s=\frac{3}{4}$($s=\frac{d}{4}$, in which $d$ denote
dimensions), this solution (Eq. (\ref{PMI3})) reduce to conformally
invariant Maxwell black hole solutions, as expected \cite{HendiES}.

In order to confirm the black hole interpretation of the solutions, we look
for the curvature singularity. To do so, we calculate the Kretschmann
scalar. Calculations show that for finite values of radial coordinate, the
Kretschmann scalar is finite. On the other hand, for very small and very
large values of $r$, we obtain
\begin{eqnarray}
\lim_{r\rightarrow 0}R_{\alpha \beta \mu \nu }R^{\alpha \beta \mu \nu }
&=&\infty ,  \label{RR0} \\
\lim_{r\rightarrow \infty }R_{\alpha \beta \mu \nu }R^{\alpha \beta \mu \nu
} &=&\frac{16\Lambda ^{2}(2\alpha ^{4}-4\alpha ^{2}+3)}{\mathcal{K}%
_{-2,1}^{2}}\left( \frac{b}{r}\right) ^{4\gamma }.  \label{RRinf}
\end{eqnarray}

\begin{figure}[tbp]
$%
\begin{array}{cc}
\epsfxsize=5cm \epsffile{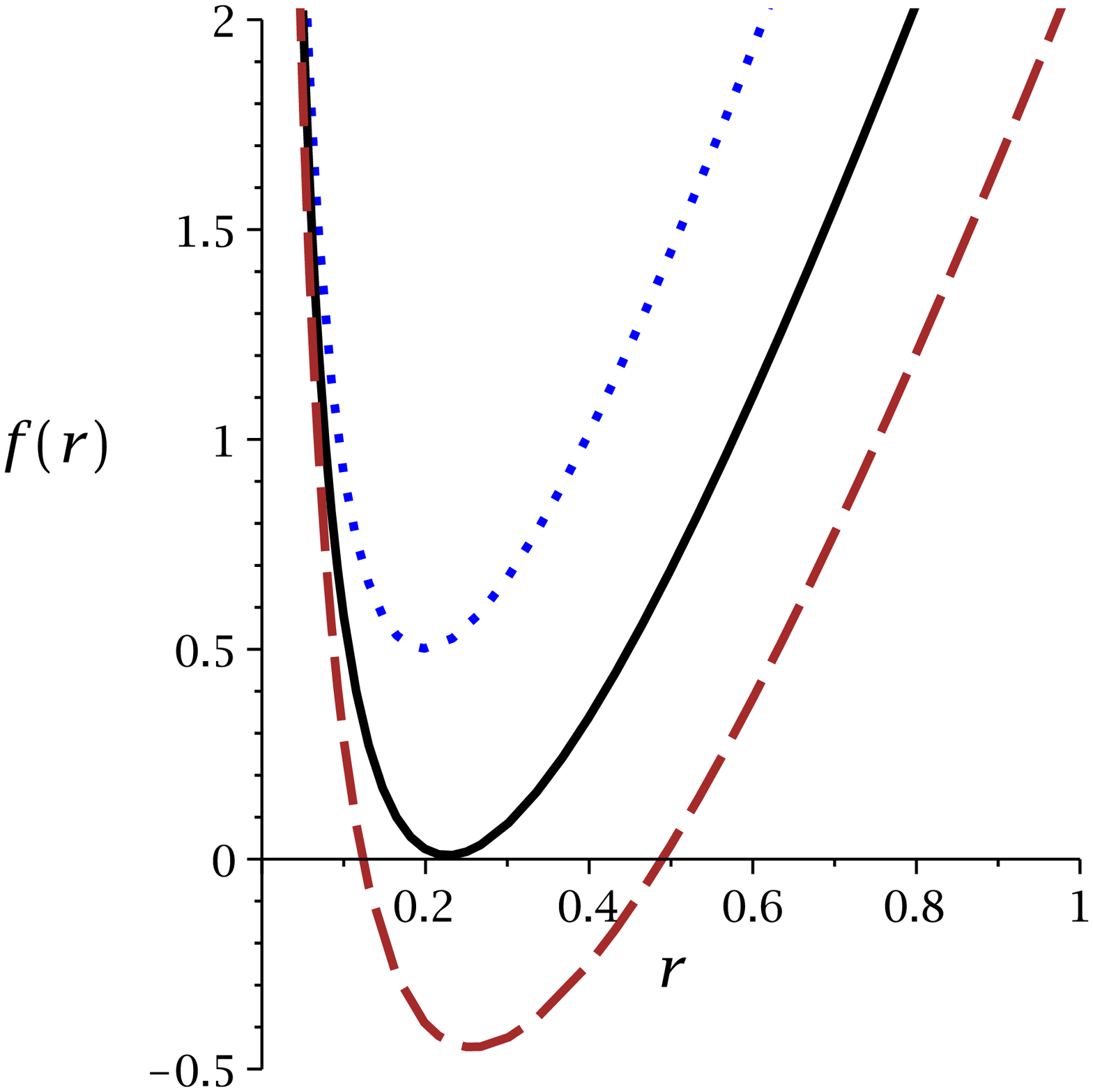} & \epsfxsize=5cm \epsffile{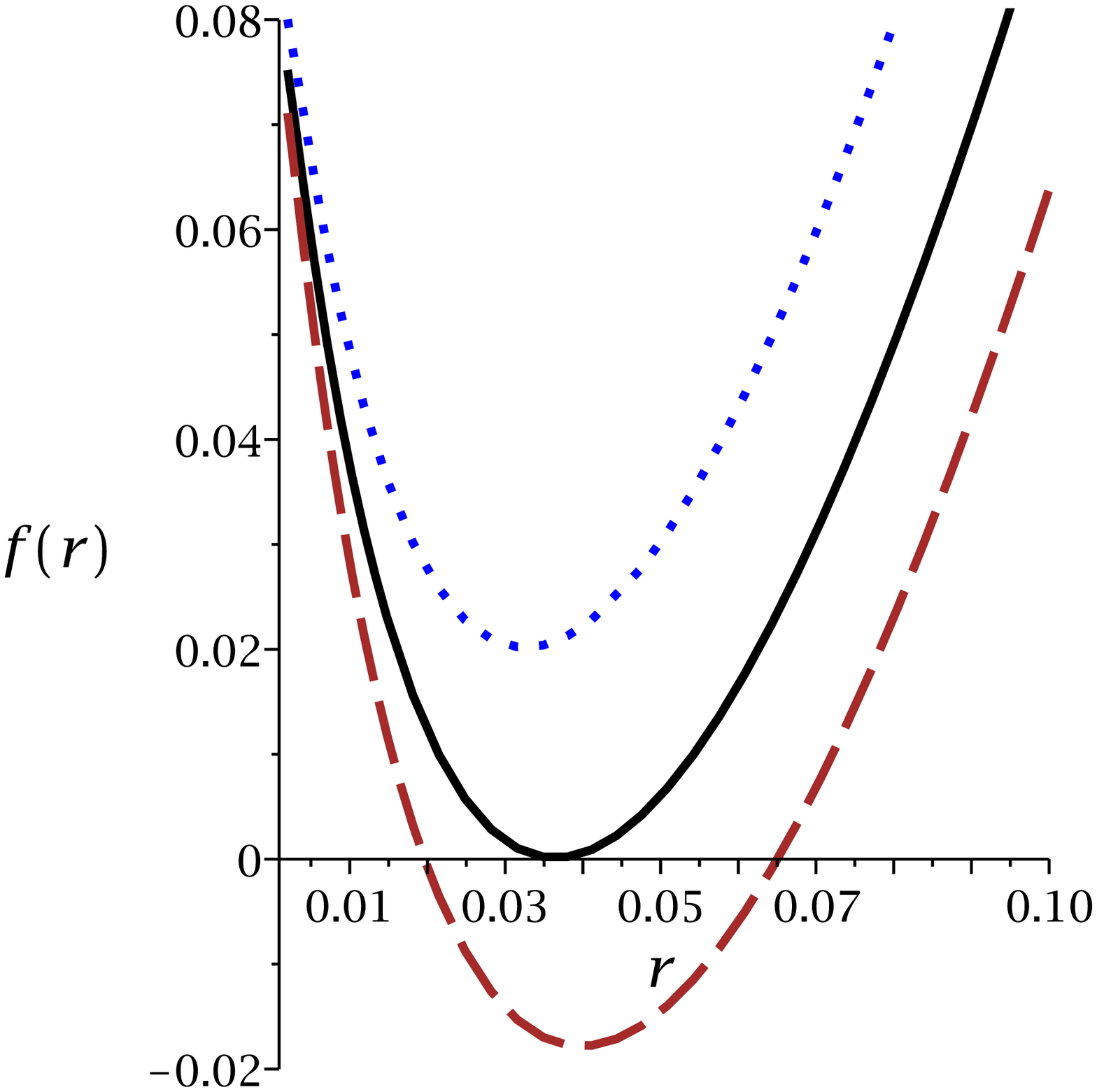}%
\end{array}
$%
\caption{$f(r)$ versus $r$ for $\Lambda =-1$, $q=0.3$, $\protect\alpha =1$
and $b=1$. \newline
left panel: for $s=0.75$, $m=4.0$ (dotted lines), $m=5.1$ (continuous line)
and $m=6.0$ (dashed line). \newline
right panel: for $s=1.25$, $m=6.0$ (dotted lines), $m=6.1$ (continuous line)
and $m=6.2$ (dashed line).}
\label{metricFig}
\end{figure}

The equation (\ref{RR0}) confirms that there is an essential singularity
located at $r=0$, while Eq. (\ref{RRinf}) shows that for nonzero $\alpha $,
the asymptotical behavior of solutions is not (A)dS. It is easy to show that
the metric function may contain real positive roots (see Fig. \ref{metricFig}%
), and therefore, the curvature singularity can be covered with an event
horizon and interpreted as a black hole.


\section{Thermodynamical Quantities}

\label{Therm}

Now, we are in a position to calculate thermodynamic and conserved
quantities of the obtained solutions and examine the validity of first law
of thermodynamics.

In order to study the temperature, we use the concept of surface gravity
which leads to
\begin{equation}
T=-\frac{\Lambda \mathcal{K}_{1,1}}{2\pi }b^{2\gamma }r_{+}^{\frac{\mathcal{K%
}_{1,-1}}{\mathcal{K}_{1,1}}}+\frac{2^{s-1}s\left( s-\frac{1}{2}\right)
q^{2s}\mathcal{K}_{1,1}}{\pi \left( s\mathcal{K}_{-1,1}-\alpha ^{2}\right)
r_{+}^{1/\left( 2s-1\right) }}.  \label{temp}
\end{equation}

On the other hand, one can use the area law for extracting modified version
of the entropy related to the Einsteinian class of black objects with the
following structure
\begin{equation}
S=\frac{\pi r_{+}}{2}\left( \frac{b}{r_{+}}\right) ^{\gamma },
\label{entropy}
\end{equation}%
in which by setting $\alpha =0$, the entropy of charged BTZ black holes is
obtained. In order to find the total electric charge of solutions, one can
use the Gauss law. Calculating the flux of electric field helps us to find
the total electric charge with the following form
\begin{equation}
Q=\frac{q^{2s-1}2^{s-2}s^{2}}{s\mathcal{K}_{-1,1}+\alpha ^{2}}.  \label{Q}
\end{equation}

Next, we are interested in obtaining the electric potential. Using following
standard relation, one can obtain the electric potential at the event
horizon with respect to the infinity as a reference
\begin{equation}
U\left( r\right) =\left. A_{\mu }\chi ^{\mu }\right\vert _{r\longrightarrow
\infty }-\left. A_{\mu }\chi ^{\mu }\right\vert _{r\longrightarrow r_{+}}=%
\frac{\left( 2s-1\right) qb^{\gamma }\mathcal{K}_{1,1}}{\mathcal{K}_{2,1}-2s}%
r_{+}^{\frac{2s-\mathcal{K}_{2,1}}{\left( 2s-1\right) \mathcal{K}_{1,1}}}.
\end{equation}

Finally, according to the definition of mass due to Abbott and Deser \cite%
{AD1,AD2,AD3}, the total mass of solution is
\begin{equation}
M=\frac{b^{\gamma }}{8\mathcal{K}_{1,1}}m.  \label{Mass}
\end{equation}

It is worthwhile to mention that for limiting case of $\alpha =0$, Eq. (\ref%
{Mass}) reduces to the mass of charged BTZ black holes \cite{Sheykhi}.

Now, we are in a position to check the validity of first law of
thermodynamics. To do so, first, we calculate the geometrical mass, $m$, by
using $f\left( r=r_{+}\right) =0$. Then by employing the obtained
geometrical mass, we rewrite the mass relation, Eq. (\ref{Mass}), in the
following form
\begin{equation}
M(r_{+},q)=\frac{b^{3\gamma }\mathcal{K}_{1,1}r_{+}^{\frac{\mathcal{K}_{2,-1}%
}{\mathcal{K}_{-2,1}}}}{4\mathcal{K}_{-2,1}}\Lambda +\frac{s\left(
2s-1\right) ^{2}\left( 2q^{2}\right) ^{s}b^{\gamma }\mathcal{K}_{1,1}r_{+}^{%
\frac{2s-\mathcal{K}_{2,1}}{\mathcal{K}_{1,1}\left( 2s-1\right) }}}{8\left( s%
\mathcal{K}_{-1,1}-\alpha ^{2}\right) \left( 2s-\mathcal{K}_{2,1}\right) }.
\label{Mass2}
\end{equation}

It is a matter of calculation to show that
\begin{equation}
\left( \frac{\partial M}{\partial S}\right) _{Q}=T\text{ \ \ \ \ }\&\text{\
\ \ \ \ }\left( \frac{\partial M}{\partial Q}\right) _{S}=U.
\end{equation}

Therefore, it is proved that the first law is valid as
\begin{equation}
dM=\left( \frac{\partial M}{\partial S}\right) _{Q}dS+\left( \frac{\partial M%
}{\partial Q}\right) _{S}dQ.
\end{equation}

\section{Thermodynamical behavior}

\label{TB}

\subsection{Mass}

In usual black holes thermodynamics, the mass of black hole is interpreted
as internal energy. In classical black hole thermodynamics, it is necessary
to have positive value for internal energy of the system, hence, the mass of
black holes. Therefore, in this section we will study the conditions which
are determining the positivity and negativity of this conserved quantity.

The mass of these black holes (\ref{Mass2}) is constructed by two terms. The
first term of Eq. (\ref{Mass2}) is related to dilatonic gravity, and the
second term includes a combination of electromagnetic and dilatonic
parameters. The positivity of first term is determined by $K_{-2,1}>0$. The
second term has two specific parts which determine the positivity and
negativity of it. These terms are $\left( s\mathcal{K}_{-1,1}-\alpha
^{2}\right) $ and $\ \left( 2s-\mathcal{K}_{2,1}\right) $. In order for the
second term has positive effects on the total mass of black holes, the
following conditions should be satisfied simultaneously

\begin{eqnarray*}
\left( s\mathcal{K}_{-1,1}-\alpha ^{2}\right) &>&0\text{ \ \ \& \ }\left( 2s-%
\mathcal{K}_{2,1}\right) >0, \\
\left( s\mathcal{K}_{-1,1}-\alpha ^{2}\right) &<&0\text{ \ \ \& \ }\left( 2s-%
\mathcal{K}_{2,1}\right) <0.
\end{eqnarray*}

The existence of mentioned conditions for positivity and negativity of the
mass indicates that under certain circumstances (suitable choices of
different parameters), it is possible for the mass of black holes to be;
completely negative/positive or a root may exist for the mass of black holes
which results into existence of two regions of negative and positive mass
for black holes. We see that the positivity/negativity of first term of Eq. (%
\ref{Mass2}) only depends on dilatonic gravity while for the second term of
Eq. (\ref{Mass2}), due to the coupling of dilatonic gravity and
electromagnetic field in the action, it depends on nonlinearity and
dilatonic parameters.

\subsection{Temperature}

Now, we focus on the behavior of temperature. Thermodynamically speaking,
for black holes, existence of negative temperature indicates non-physical
systems, while, positive temperature is denoted as systems being physical.
In other words, negativity and positivity of the temperature confirm whether
obtained solutions for black holes are physical or non-physical ones. Here,
the temperature of these black holes (\ref{temp}) contains two terms; $%
\Lambda $ term which is purely due to the dilatonic part of action and
electric charge term which is a combination of dilatonic parameter and
electric charge.

The total contribution of $\Lambda$ term depends on the choices of $\Lambda $
itself. If $\Lambda >0$, the contribution of this term will be towards the
negativity whereas for $\Lambda <0$, this term will have positive effects on
values of temperature. As for the charge term, its effects are determined by
two factors which are $\left( s-\frac{1}{2}\right) $ and $\left( s\mathcal{K}%
_{-1,1}-\alpha ^{2}\right) $. Considering our earlier discussions, ($s>\frac{%
1}{2}$), one can automatically conclude that the contributions of this term
is positive. Therefore, the negativity or positivity of charge term only
depends on the sign of $\left( s\mathcal{K}_{-1,1}-\alpha ^{2}\right) $.
Meaning that for
\[
s\mathcal{K}_{-1,1}-\alpha ^{2}>0,
\]%
the charge term will be positive and its effects are toward positivity of
the temperature, whereas the opposite is observed for violation of this
condition. The above condition presents the direct interaction of
electromagnetic and dilatonic fields in an elegant way. We see that the
coupling of electromagnetic field and dilaton gravity results into specific
contribution in thermodynamical structure of the solutions and presents a
tool for tuning the effects of dilaton gravity and electromagnetic field for
specific purposes. Returning to temperature, interestingly, we notice that
in $\Lambda $ term, the power of event horizon only depends on dilatonic
parameter whereas in charge term, it only depends on nonlinearity parameter,
$s$. Here, we see that regardless of the coupling between electromagnetic
field and dilatonic gravity, the power of horizon radius has been separated
into two groups depending on different parts of the action.
Thermodynamically speaking, the effectiveness of horizon radius has been
separated into two branches which depends on the choices of dilatonic and
nonlinearity parameters. The structure of action introduced such property
into thermodynamical structure of the black holes. Therefore, it is possible
to modify the effect of horizon radius in one branch (whether
electromagnetic part or dilatonic gravity) without any concern regarding the
other part.

\subsection{Heat capacity}

Now, we study the positivity/negativity of heat capacity. The heat capacity
has specific information regarding the stability/instability of system and
phase transitions. The phase transitions usually take place when system is
in an unstable state. In other words, unstable systems go under a phase
transition to acquire stability. The phase transition in heat capacity are
recognized when the heat capacity meets a discontinuity, hence divergency.
In other words, divergencies of the heat capacity are where systems go under
phase transitions. The thermal stability/instability of the system is
determined by the sign of heat capacity; the negative values are
representing the system being in unstable state while the positivity of the
heat capacity is denoted as system being thermally stable.

The heat capacity is given by
\begin{equation}
C_{Q}=\left( \frac{\partial M}{\partial S}\right) _{Q}\left( \frac{\partial
^{2}M}{\partial S^{2}}\right) _{Q}^{-1},
\end{equation}%
where by using the first law of black holes thermodynamics, one can rewrite
it as

\begin{equation}
C_{Q}=T\frac{\left( \frac{\partial S}{\partial r_{+}}\right) _{Q}}{\left(
\frac{\partial T}{\partial r_{+}}\right) _{Q}}.  \label{heat}
\end{equation}

Using this relation, it is possible to extract two thermodynamical important
points; bound and phase transition points.

\begin{equation}
\left\{
\begin{array}{cc}
T=0 & bounded\text{ }point \\
&  \\
\left( \frac{\partial T}{\partial r_{+}}\right) _{Q}=0 & phase\text{ }%
transition\text{ }point%
\end{array}%
\right. .  \label{phase}
\end{equation}

The bound point is where the heat capacity, hence temperature, meets a root.
Since this point separates physical (positive temperature) form non-physical
(negative temperature) solutions, it is called bound point. Using the
obtained temperature (\ref{temp}) and entropy (\ref{entropy}), it is a
matter of calculation to show that the heat capacity for these black holes
is obtained as
\begin{equation}
C_{Q}=\frac{\pi \left[ \Lambda \left( \alpha ^{2}-s\mathcal{K}_{-1,1}\right)
b^{2\gamma }r_{+}^{\left( 2s\mathcal{K}_{3,1}+\mathcal{K}_{-2,1}\right)
}+s\left( 2q^{2}\right) ^{s}\left( s-\frac{1}{2}\right) r_{+}^{\left( 4s%
\mathcal{K}_{1,1}-\mathcal{K}_{2,1}\right) }\right] }{2r_{+}\left[ \Lambda
\mathcal{K}_{-1,1}\left( s\mathcal{K}_{-1,1}-\alpha ^{2}\right) b^{\gamma
}r_{+}^{2\left( s+\alpha ^{2}\right) }-\frac{s\left( 2q^{2}\right) ^{s}%
\mathcal{K}_{1,1}}{2b^{\gamma }}r_{+}^{2s\alpha ^{2}}\right] }.  \nonumber
\end{equation}

The numerator and denominator of heat capacity consist of $\Lambda $,
dilatonic parameter and electric charge. For having thermally stable
solutions, hence positive heat capacity, two different set of conditions
exist which must be satisfied. These conditions are

\[
\Lambda \left( \alpha ^{2}-s\mathcal{K}_{-1,1}\right) >0\text{ \ \ \& \ \ }%
\Lambda \mathcal{K}_{-1,1}\left( s\mathcal{K}_{-1,1}-\alpha ^{2}\right)
b^{\gamma }r_{+}^{2\left( s+\alpha ^{2}\right) }-\frac{s\left( 2q^{2}\right)
^{s}\mathcal{K}_{1,1}}{2b^{\gamma }}r_{+}^{2s\alpha ^{2}}>0,
\]

\ \

\[
\frac{\Lambda \left( \alpha ^{2}-s\mathcal{K}_{-1,1}\right) b^{2\gamma }}{%
r_{+}^{-\left( 2s\mathcal{K}_{3,1}+\mathcal{K}_{-2,1}\right) }}+\frac{%
s\left( 2q^{2}\right) ^{s}\left( s-\frac{1}{2}\right) }{r_{+}^{\left(
\mathcal{K}_{2,1}-4s\mathcal{K}_{1,1}\right) }}<0\ \ \ \text{\&}\ \ \ \frac{%
\Lambda \mathcal{K}_{-1,1}\left( s\mathcal{K}_{-1,1}-\alpha ^{2}\right)
b^{\gamma }}{r_{+}^{-2\left( s+\alpha ^{2}\right) }}-\frac{s\left(
2q^{2}\right) ^{s}\mathcal{K}_{1,1}}{2b^{\gamma }r_{+}^{-2s\alpha ^{2}}}<0.
\]

Respectively, the root (bound point) and divergence points (phase
transition) of these black holes are given by
\begin{equation}
r_{root}=\left( \frac{\Lambda b^{2\gamma }\left( s\mathcal{K}_{-1,1}-\alpha
^{2}\right) }{s\left( 2q^{2}\right) ^{s}\left( s-\frac{1}{2}\right) }\right)
^{\frac{\mathcal{K}_{1,1}\left( 2s-1\right) }{2\left( s\mathcal{K}%
_{-1,1}-\alpha ^{2}\right) }},
\end{equation}%
\begin{equation}
r_{phase~transition}=\left( \frac{2\Lambda b^{2\gamma }\mathcal{K}%
_{-1,1}\left( s\mathcal{K}_{-1,1}-\alpha ^{2}\right) }{s\left( 2q^{2}\right)
^{s}\mathcal{K}_{1,1}}\right) ^{\frac{\mathcal{K}_{1,1}\left( 2s-1\right) }{%
2\left( s\mathcal{K}_{-1,1}-\alpha ^{2}\right) }}.
\end{equation}

If $\frac{\mathcal{K}_{1,1}\left( 2s-1\right) }{2\left( s\mathcal{K}%
_{-1,1}-\alpha ^{2}\right) }>1$, then root and divergence points of the heat
capacity are decreasing functions of the electric charge and increasing
functions of $\Lambda $. If $\frac{\mathcal{K}_{1,1}\left( 2s-1\right) }{%
2\left( s\mathcal{K}_{-1,1}-\alpha ^{2}\right) }$ is even, then the
existence of positive real valued root and phase transition point for these
black holes is restricted to the following inequalities being satisfied

\[
\Lambda \left( s\mathcal{K}_{-1,1}-\alpha ^{2}\right) >0,\text{ \ \ \ \ \&\
\ \ \ }\Lambda \mathcal{K}_{-1,1}\left( s\mathcal{K}_{-1,1}-\alpha
^{2}\right) >0,
\]%
which once more, highlight the interaction between dilaton and
electromagnetic fields.

\begin{figure}[tbp]
$%
\begin{array}{cc}
\epsfxsize=5cm \epsffile{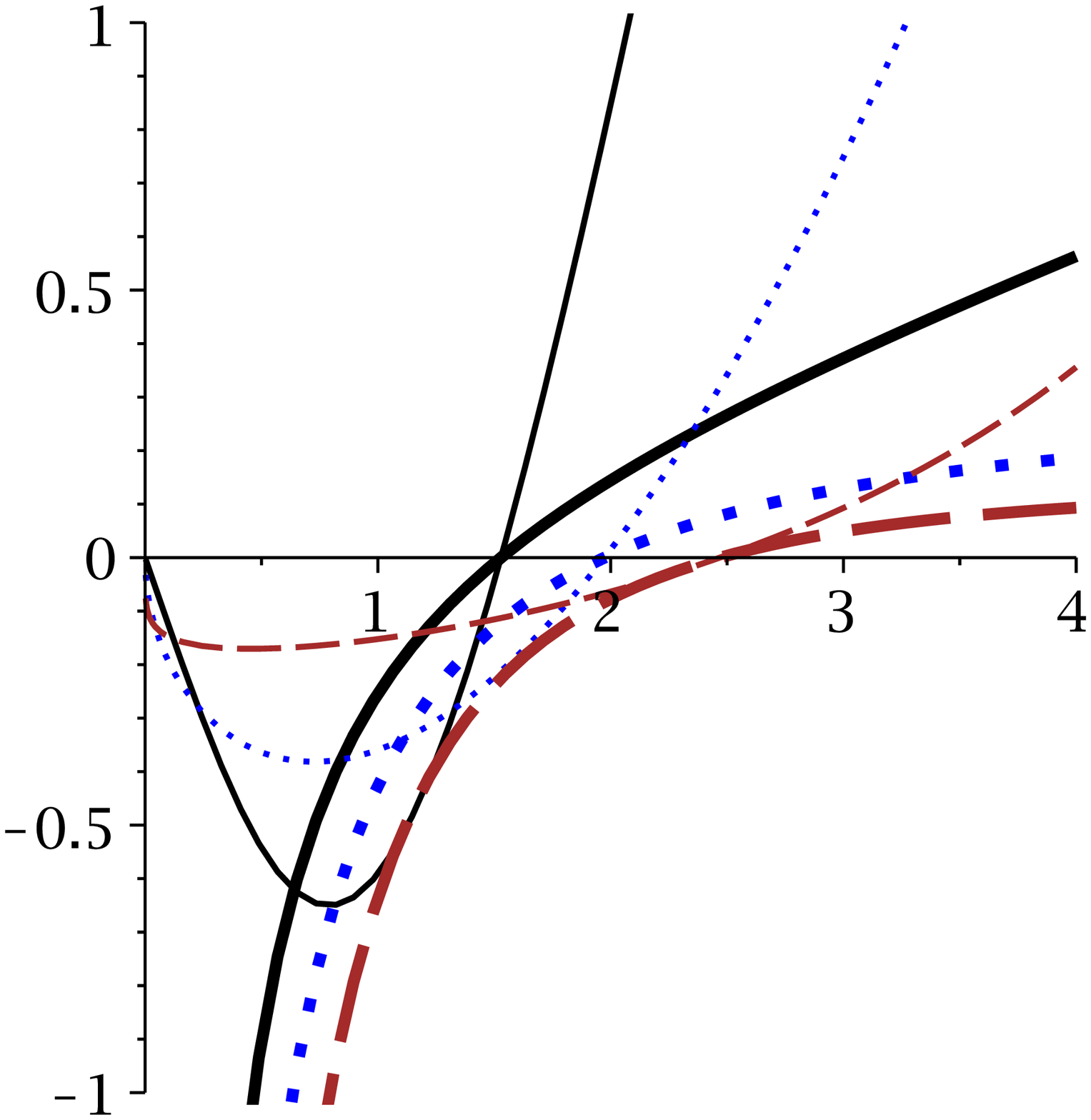} & \epsfxsize=5cm %
\epsffile{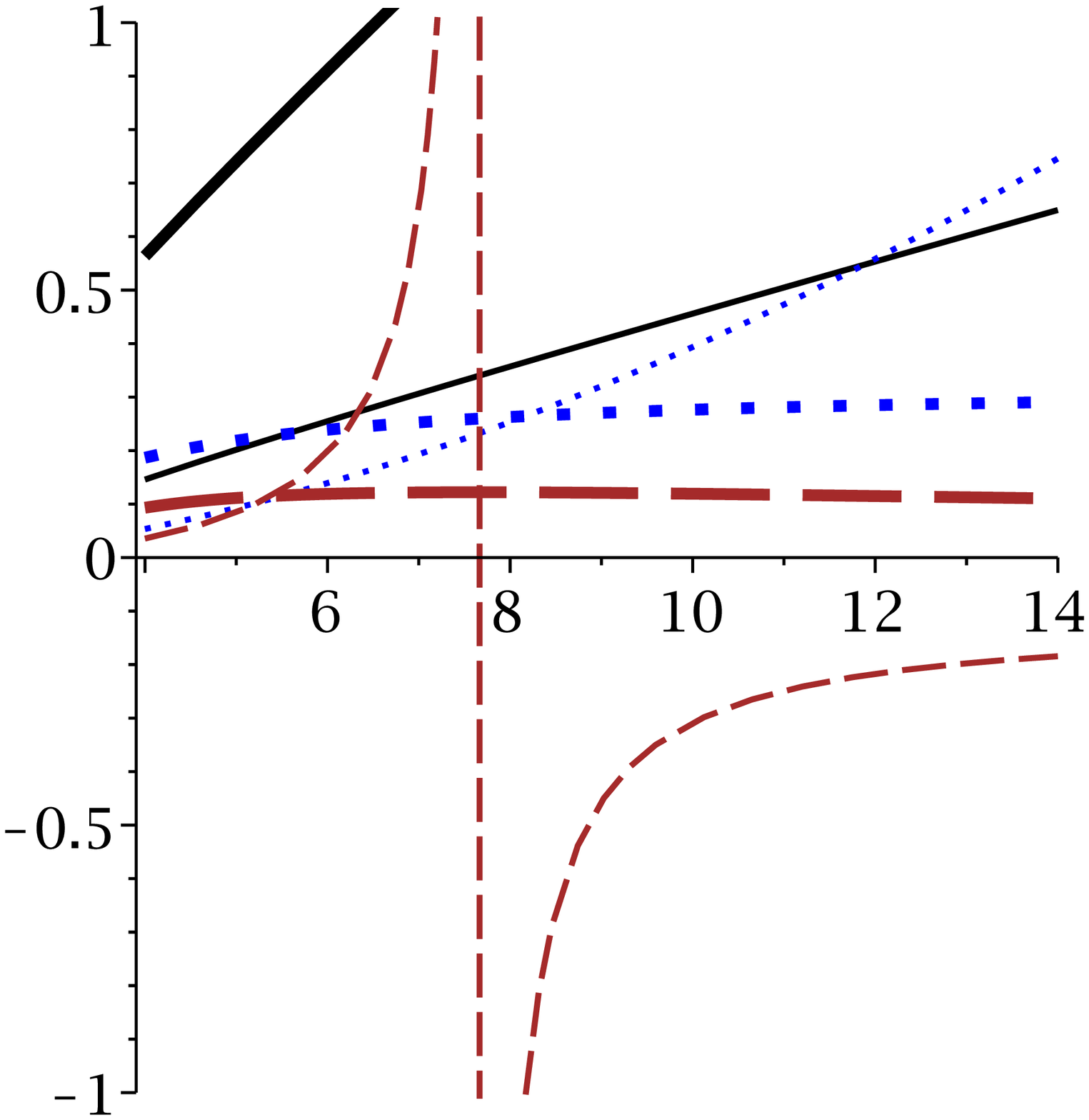} \\
\epsfxsize=5cm \epsffile{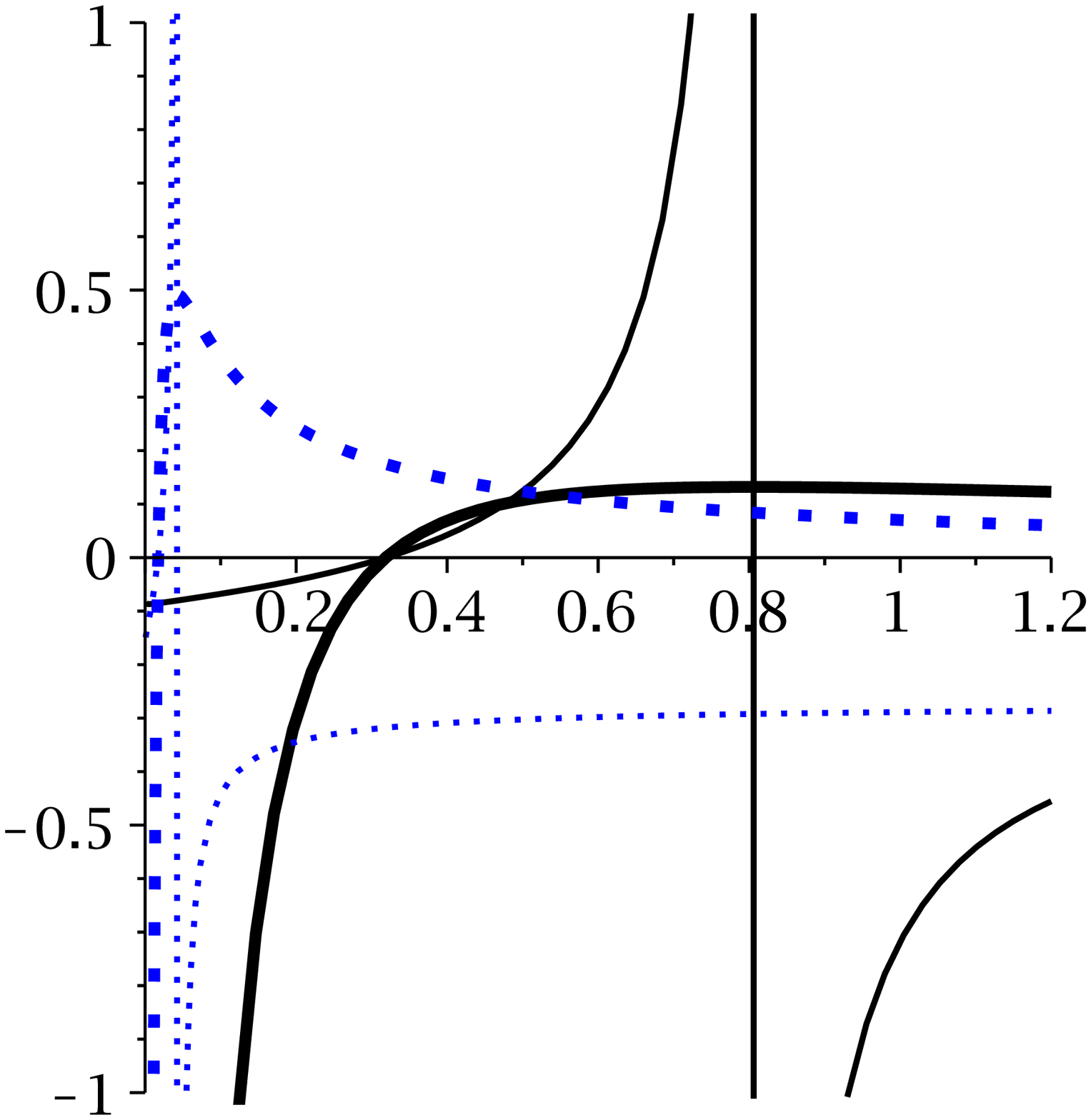} & \epsfxsize=5cm %
\epsffile{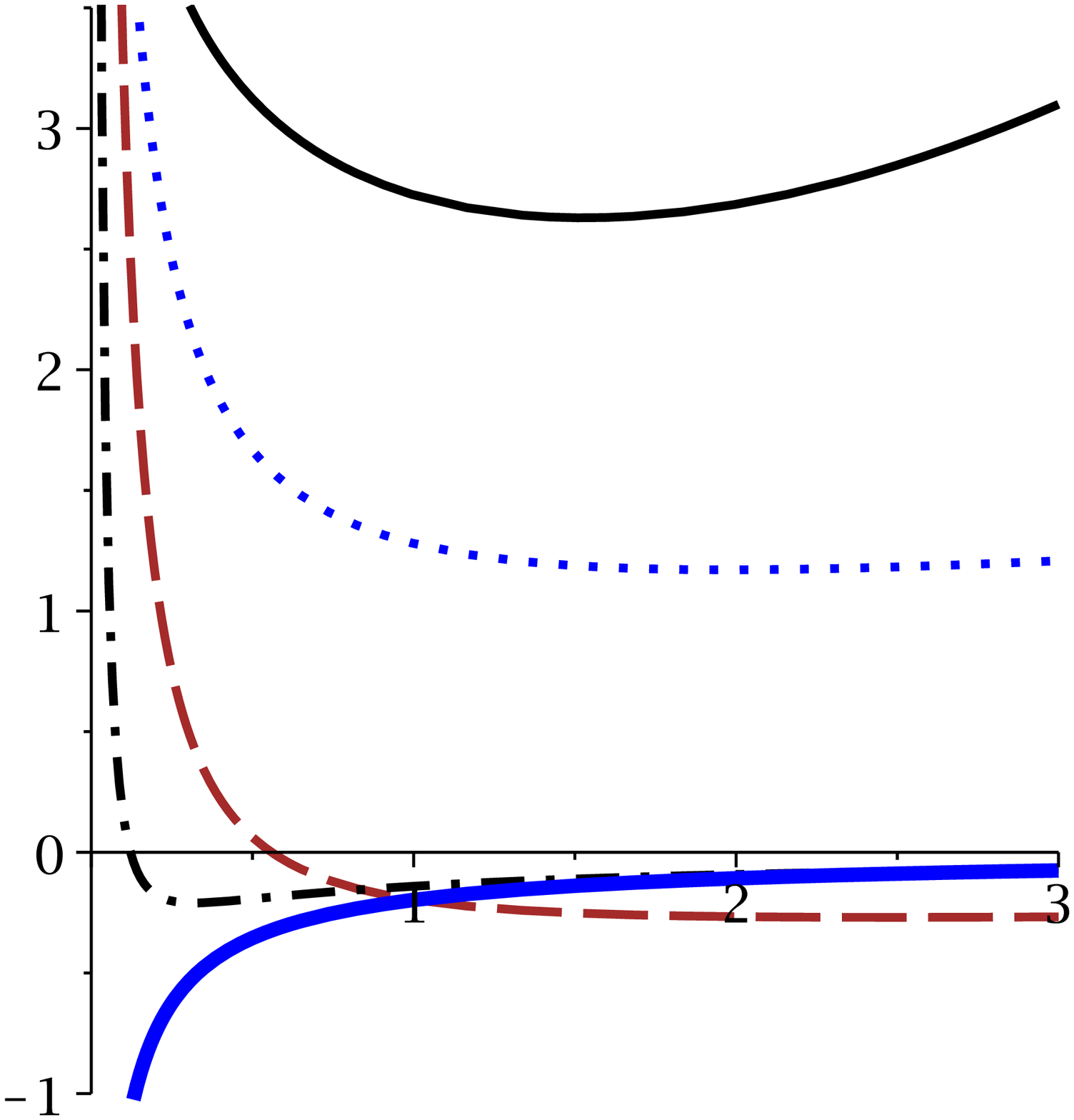}%
\end{array}
$%
\caption{For $\Lambda =-1$, $q=2$, $s=0.9$ and $b=1$. \newline
up panels: $C_{Q}$ and $T$ (bold lines) versus $r_{+}$ diagrams for $\protect%
\alpha=0$ (continuous line), $\protect\alpha=1$ (dotted line) and $\protect%
\alpha=2$ (dashed line). \newline
down left panel: $C_{Q}$ and $T$ (bold lines) versus $r_{+}$ diagrams for $%
\protect\alpha=5$ (continuous line) and $\protect\alpha=8$ (dotted line).
\newline
down right panel: $M$ versus $r_{+}$ diagrams for $\protect\alpha=0$
(continuous line), $\protect\alpha=1$ (dotted line), $\protect\alpha=2$
(dashed line), $\protect\alpha=5$ (dashed-dotted line) and $\protect\alpha=8$
(bold continuous line). }
\label{Fig1}
\end{figure}

\begin{figure}[tbp]
$%
\begin{array}{cc}
\epsfxsize=5cm \epsffile{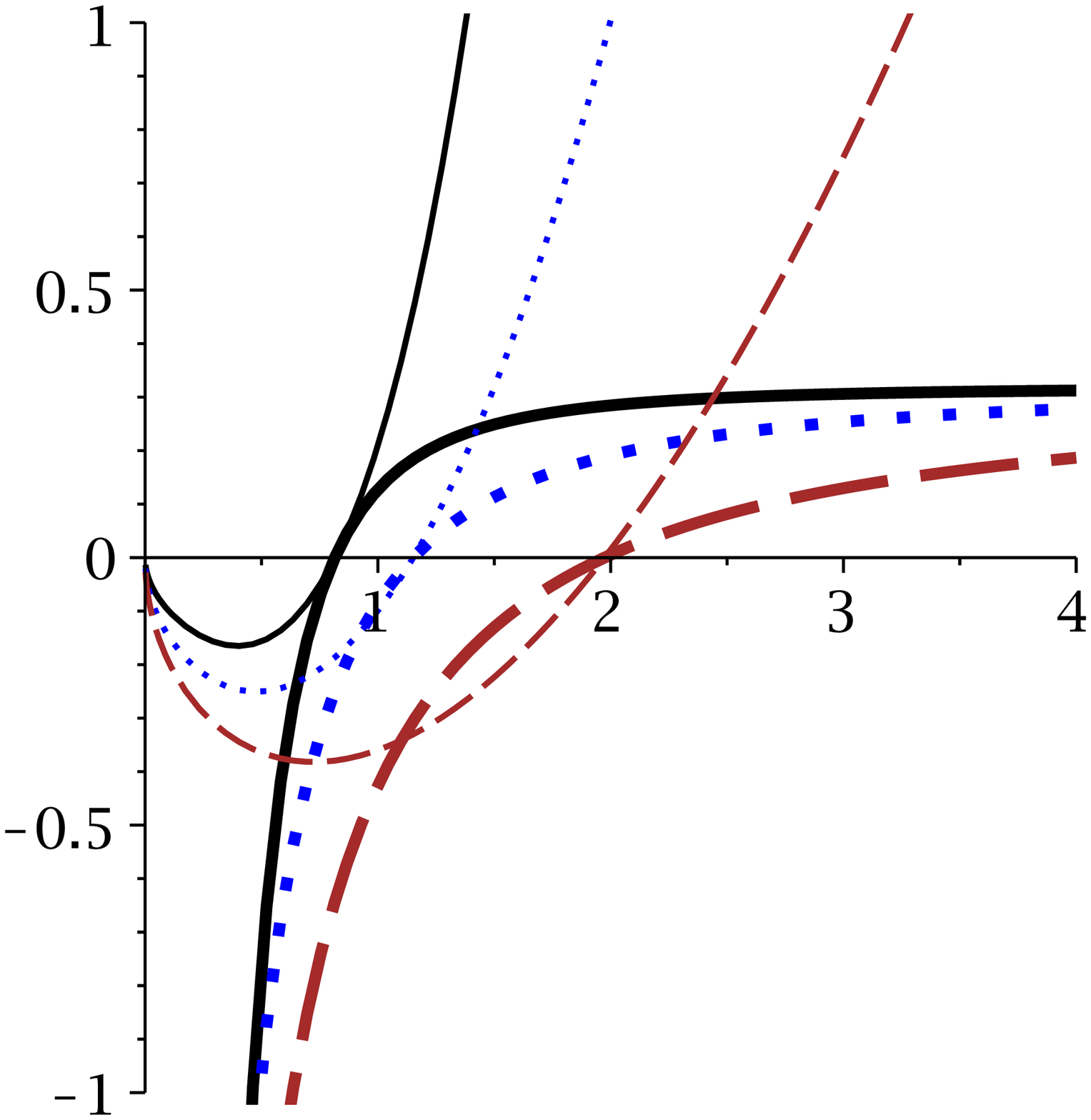} & \epsfxsize=5cm \epsffile{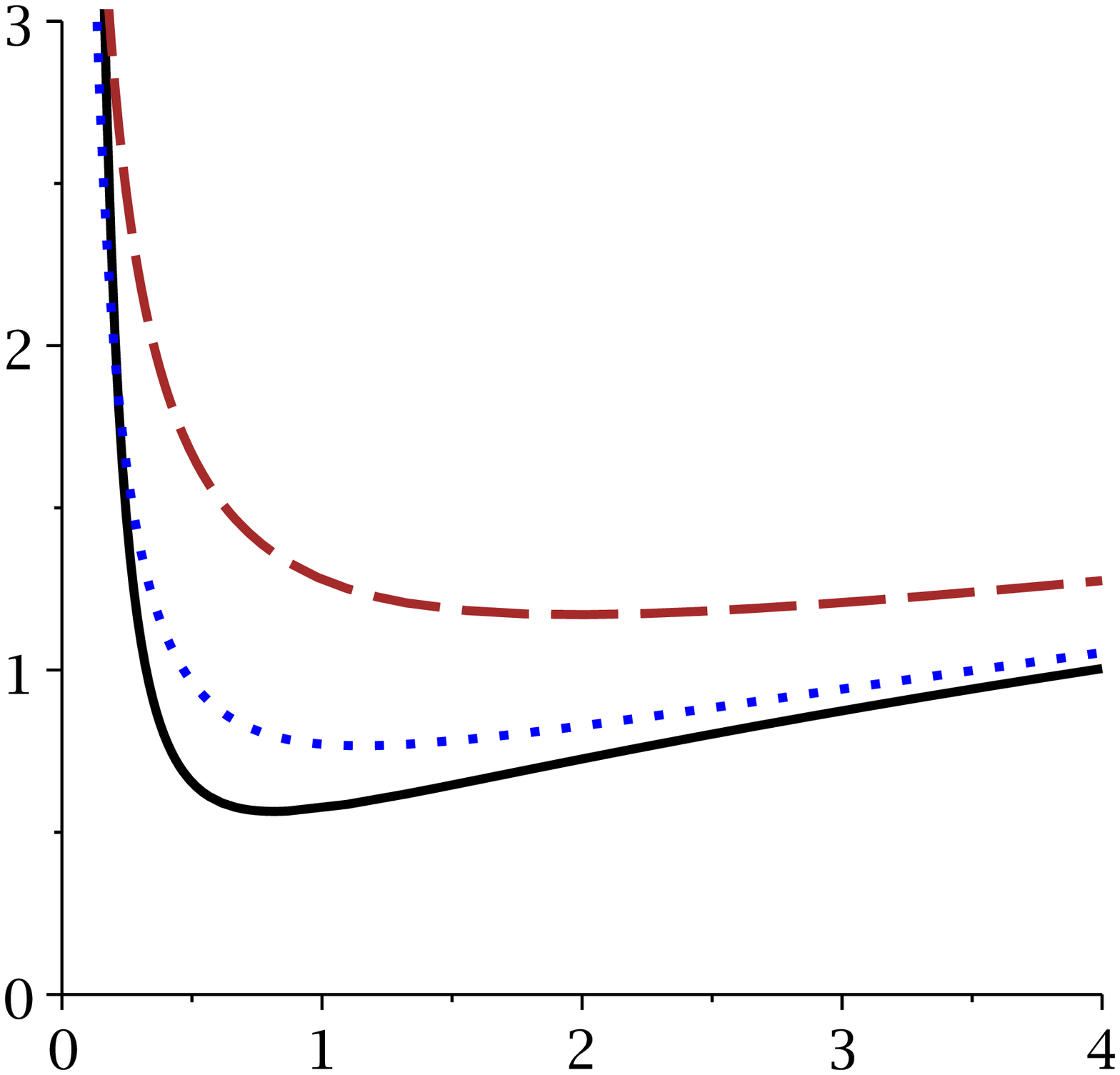}%
\end{array}
$%
\caption{For $\Lambda =-1$, $q=2$, $\protect\alpha=1$ and $b=1$; $s=0.7$
(continuous line), $s=0.8$ (dotted line) and $s=0.9$ (dashed line) \newline
left panel: $C_{Q}$ and $T$ (bold lines) versus $r_{+}$ diagrams. \newline
right panel: $M$ versus $r_{+}$ diagrams.}
\label{Fig2}
\end{figure}

\begin{figure}[tbp]
$%
\begin{array}{cc}
\epsfxsize=5cm \epsffile{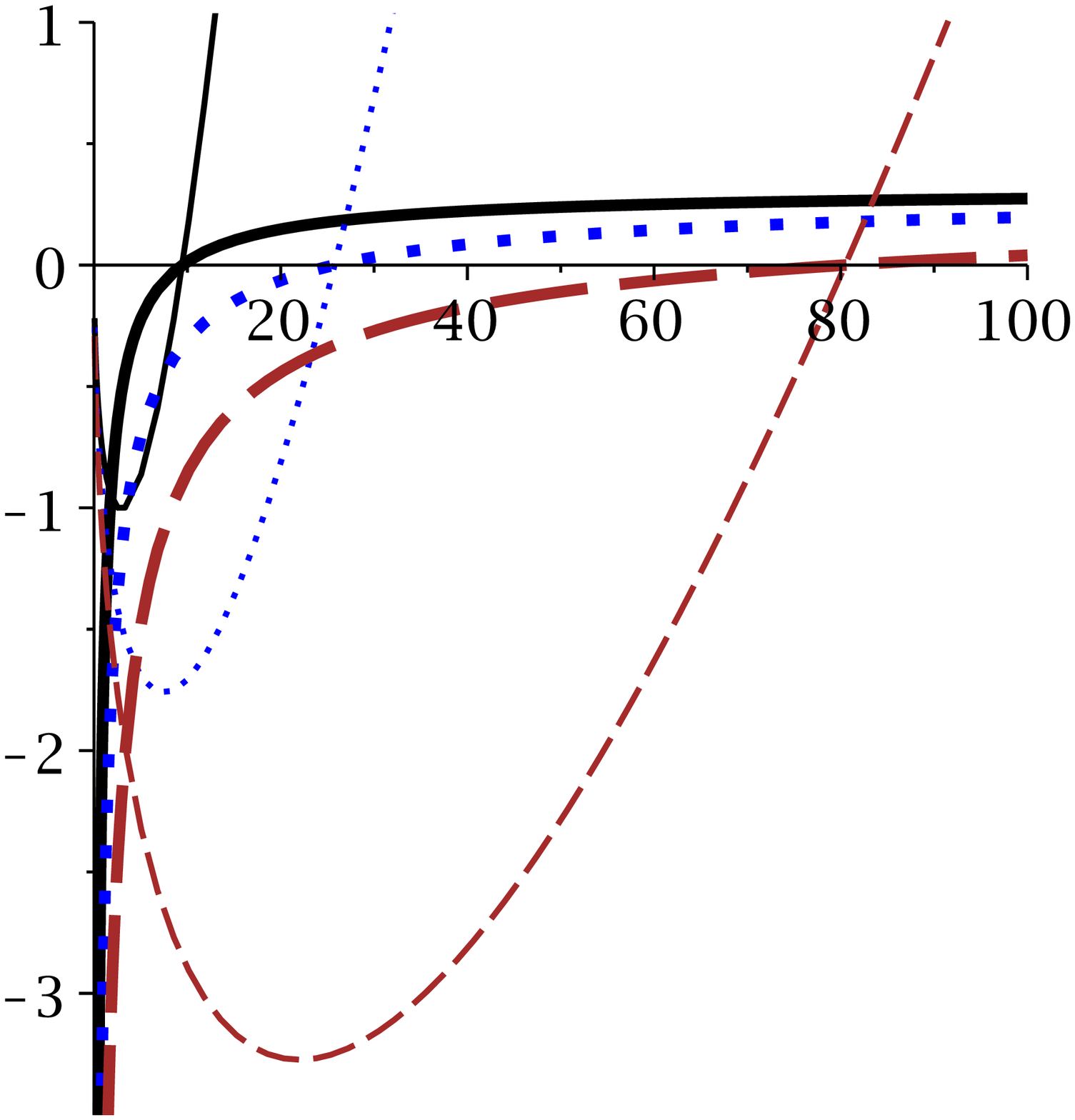} & \epsfxsize=5cm \epsffile{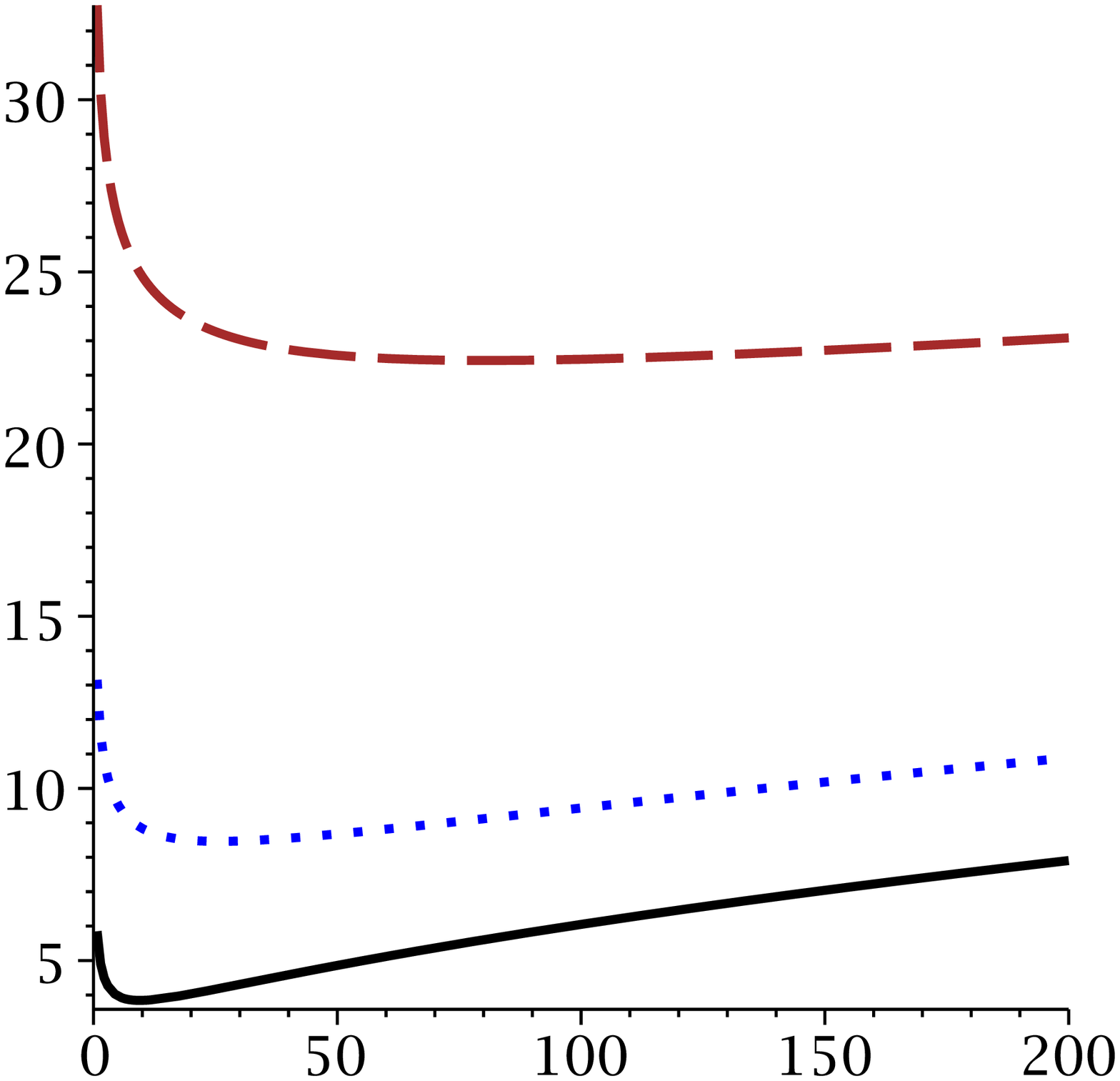}%
\end{array}
$%
\caption{For $\Lambda =-1$, $q=2$, $\protect\alpha=1$ and $b=1$; $s=1.1$
(continuous line), $s=1.2$ (dotted line) and $s=1.3$ (dashed line) \newline
left panel: $C_{Q}$ and $T$ (bold lines) versus $r_{+}$ diagrams. \newline
right panel: $M$ versus $r_{+}$ diagrams. }
\label{Fig3}
\end{figure}

\begin{figure}[tbp]
$%
\begin{array}{cc}
\epsfxsize=5cm \epsffile{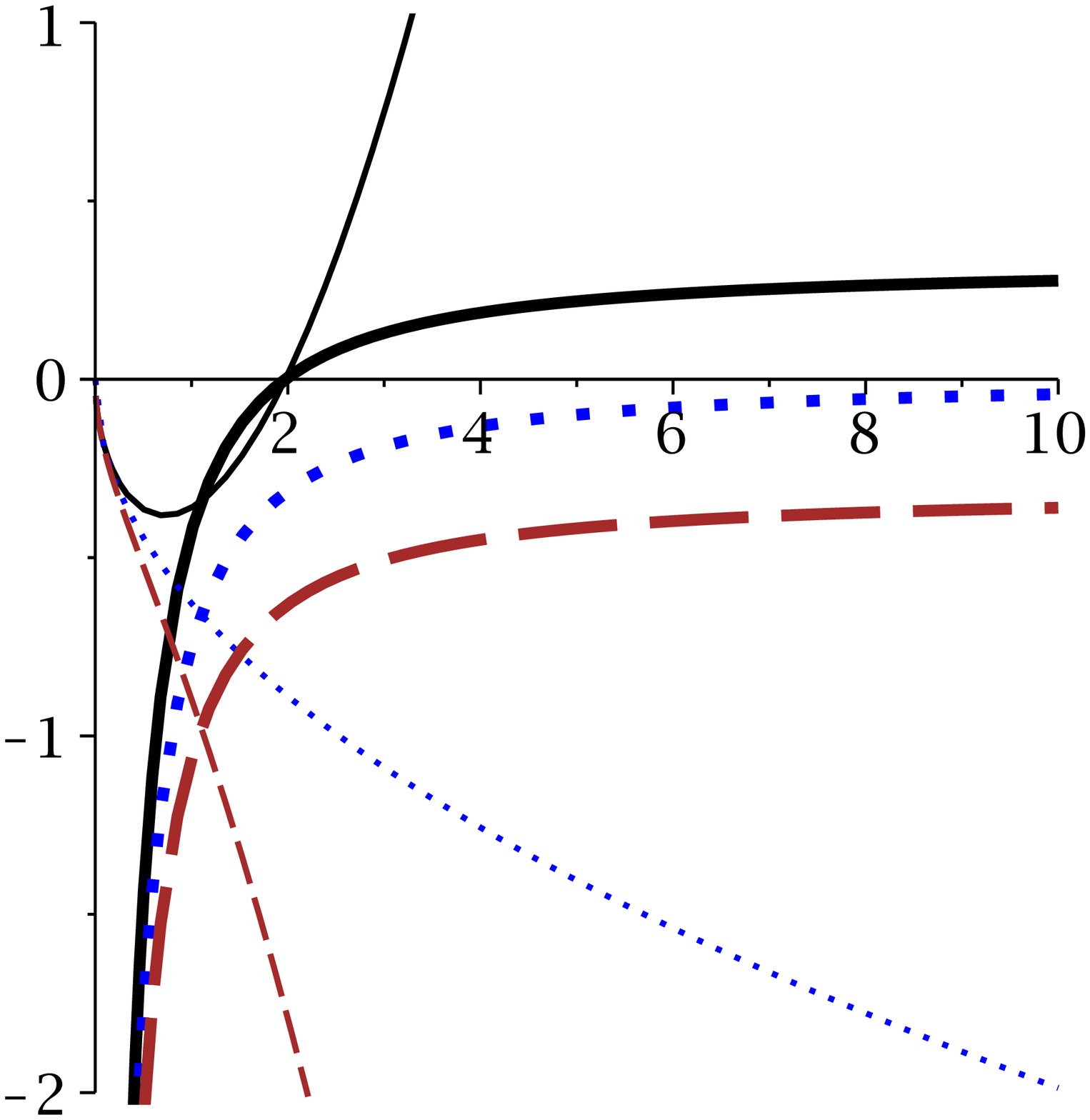} & \epsfxsize=5cm %
\epsffile{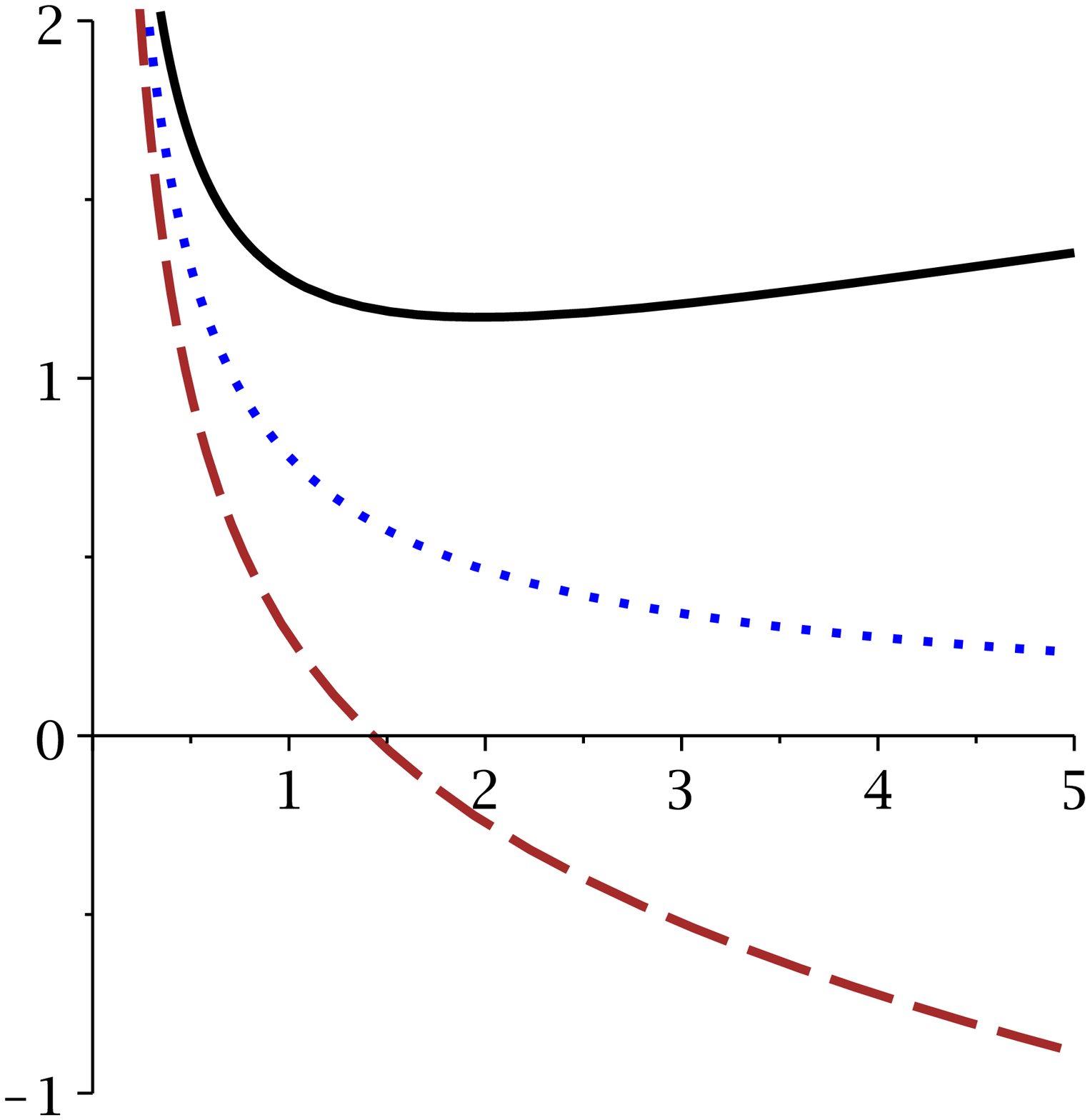}%
\end{array}
$%
\caption{For $s =0.9$, $q=2$, $\protect\alpha=1$ and $b=1$; $\Lambda=-1$
(continuous line), $\Lambda=0$ (dotted line) and $\Lambda=1$ (dashed line)
\newline
left panel: $C_{Q}$ and $T$ (bold lines) versus $r_{+}$ diagrams. \newline
right panel: $M$ versus $r_{+}$ diagrams. }
\label{Fig4}
\end{figure}

In order to elaborate the effects of different parameters on thermodynamical
behavior of the mass, temperature and heat capacity, we present a table
(table \ref{tab1}) and some diagrams (Figs. \ref{Fig1}-\ref{Fig4}).

\begin{table}[tbp]
\caption{Roots of mass ($r_{M}$), temperature ($r_{T}$) and denominator of
the heat capacity ($r_{\left( \frac{\partial T}{\partial r_{+}}\right) }$)
for $q=2$ and $b=1$. }
\label{tab1}
\begin{center}
\begin{tabular}{c|c|c|c|c|c}
\hline\hline
$s$ & $\alpha $ & $\Lambda $ & $r_{M}$ & $r_{T}$ & $r_{\left( \frac{\partial
T}{\partial r_{+}}\right) }$ \\ \hline\hline
$0.9$ & $0$ & $-1$ & $none$ & $1.528881$ & $none$ \\ \hline
$0.9$ & $1$ & $-1$ & $none$ & $1.973634$ & $none$ \\ \hline
$0.9$ & $2$ & $-1$ & $0.559382$ & $2.469012$ & $7.636975$ \\ \hline
$0.9$ & $5$ & $-1$ & $0.121752$ & $0.318611$ & $0.805420$ \\ \hline
$0.9$ & $8$ & $-1$ & $none$ & $0.017365$ & $0.042973$ \\ \hline
$0.7$ & $1$ & $-1$ & $none$ & $0.815298$ & $none$ \\ \hline
$0.8$ & $1$ & $-1$ & $none$ & $1.152417$ & $none$ \\ \hline
$1.1$ & $1$ & $-1$ & $none$ & $9.452169$ & $none$ \\ \hline
$1.2$ & $1$ & $-1$ & $none$ & $25.77402$ & $none$ \\ \hline
$1.3$ & $1$ & $-1$ & $none$ & $80.47861$ & $none$ \\ \hline
$0.9$ & $1$ & $0$ & $none$ & $none$ & $none$ \\ \hline
$0.9$ & $1$ & $1$ & $1.426900$ & $none$ & $none$ \\ \hline
\end{tabular}%
\\[0pt]
\end{center}
\end{table}

Evidently, depending on the choices of different parameters, these black
holes may enjoy a phase transition in their phase diagrams. This phase
transition is between large unstable black holes to small stable ones (see
up panel of Fig. \ref{Fig1}). The region of stability for this case is
located between a bound and divergence point, whereas after the divergency,
black holes are within physical region (positive temperature and mass) but
are unstable (see up panel of Fig. \ref{Fig1}). In another case, only a
bound point exists where the physical stable solutions are observed after
the bound point.

The critical horizon radius in which phase transition takes place is a
decreasing function of the dilatonic parameter. Whereas, the bound point
depending on the range of dilatonic parameter, could be increasing or
decreasing function of dilatonic parameter (see table \ref{tab1} and also
Fig. \ref{Fig1}). On the other hand, the bound point is an increasing
function of the nonlinearity parameter (see Figs. \ref{Fig2} and \ref{Fig3}%
). Also, there is one bound point for the negative values of cosmological
constant and for other values of the cosmological constant (positive and
zero) the black holes are not within physical region (see Fig. \ref{Fig4}).
By taking a closer look at the table \ref{tab1}, one can notice that it is
possible to obtain root for mass as well. This indicates that there exists a
region where internal energy of the system is negative. This shows that the
behavior of mass of the black hole imposes specific restrictions for
solutions being physical/non-physical ones.

For the past few years, the studies that were conducted regarding
thermodynamics of the black holes, have neglected the mass of black hole and
its thermodynamical behavior. Here, we have shown that thermodynamical
behavior of this quantity and its positivity/negativity play crucial roles
for studying thermodynamics of the black holes. In fact, without having a
well defined and positive internal energy (mass) for black holes and
considering its behavior, assumptions and physical statements regarding to
thermodynamical structure of the black holes will not be complete and even
in some cases it could be wrong and misleading. The results that are derived
for quantities such as temperature, heat capacity and etc. should be
evaluated by considering the behavior of mass as well. Remembering that some
important applications of the black holes and their thermodynamics, lie
within AdS/CFT correspondence, the necessity of studying the mass of black
holes in more details could be highlighted.

\section{Geometrical thermodynamics}

\label{GT}

In this section, we employ geometrical thermodynamic approach to investigate
thermodynamical properties of the black holes. The geometrical thermodynamic
method provides the possibility of studying thermodynamical properties of
the system through Riemannian geometry. In this method, one constructs
thermodynamical phase space of the system and use its Ricci scalar to
investigate thermodynamical properties. The divergencies of Ricci scalar in
the phase space marks two important points: bound and phase transition
points. In other words, divergencies (phase transition points) and bound
points of the thermodynamical systems coincide with the divergencies of
Ricci scalar. There are several approaches for constructing thermodynamical
phase space which among them one can name; Weinhold \cite%
{WeinholdI,WeinholdII}, Ruppeiner \cite{RuppeinerI,RuppeinerII}, Quevedo
\cite{QuevedoI,QuevedoII} and HPEM \cite{HPEMI,HPEMII,HPEMIII}. Recent
studies in the context of black holes thermodynamics revealed that Weinhold,
Ruppeiner and Quevedo metrics may lead to inconsistent results regarding
thermodynamical behavior of the system \cite{HPEMI,HPEMII,HPEMIII}. In other
words, cases of mismatch between the divergencies of Ricci scalar and, bound
and phase transition points or existence of extra divergencies unrelated to
mentioned points were reported. To overcome the shortcomings of mentioned
methods, HPEM formulation was introduced and it was shown that specific
structure of this metric provides satisfactory results regarding the
geometrical thermodynamics of different classes of black holes.

\begin{figure}[tbp]
$%
\begin{array}{ccc}
\epsfxsize=5cm \epsffile{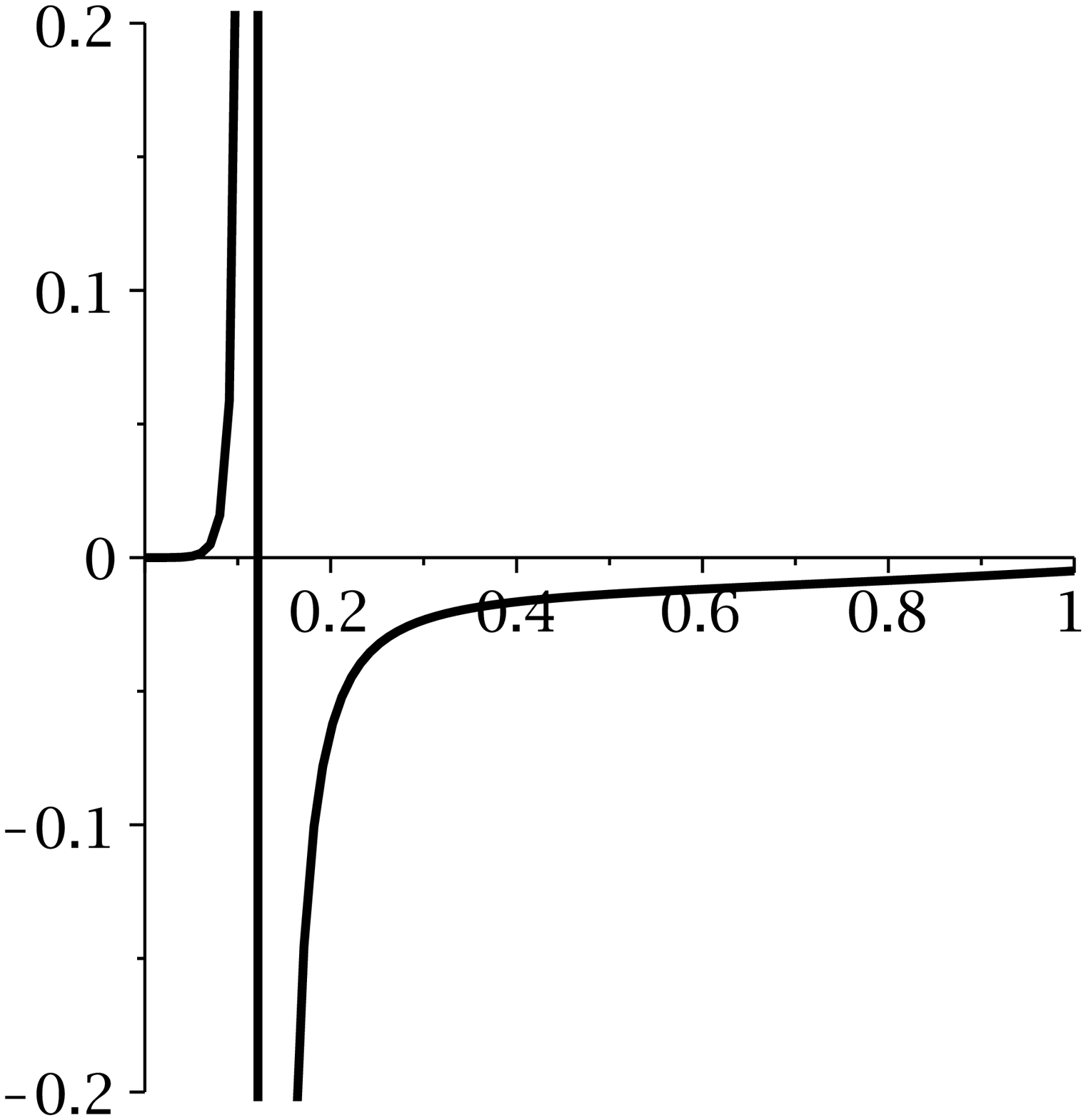} & \epsfxsize=5cm %
\epsffile{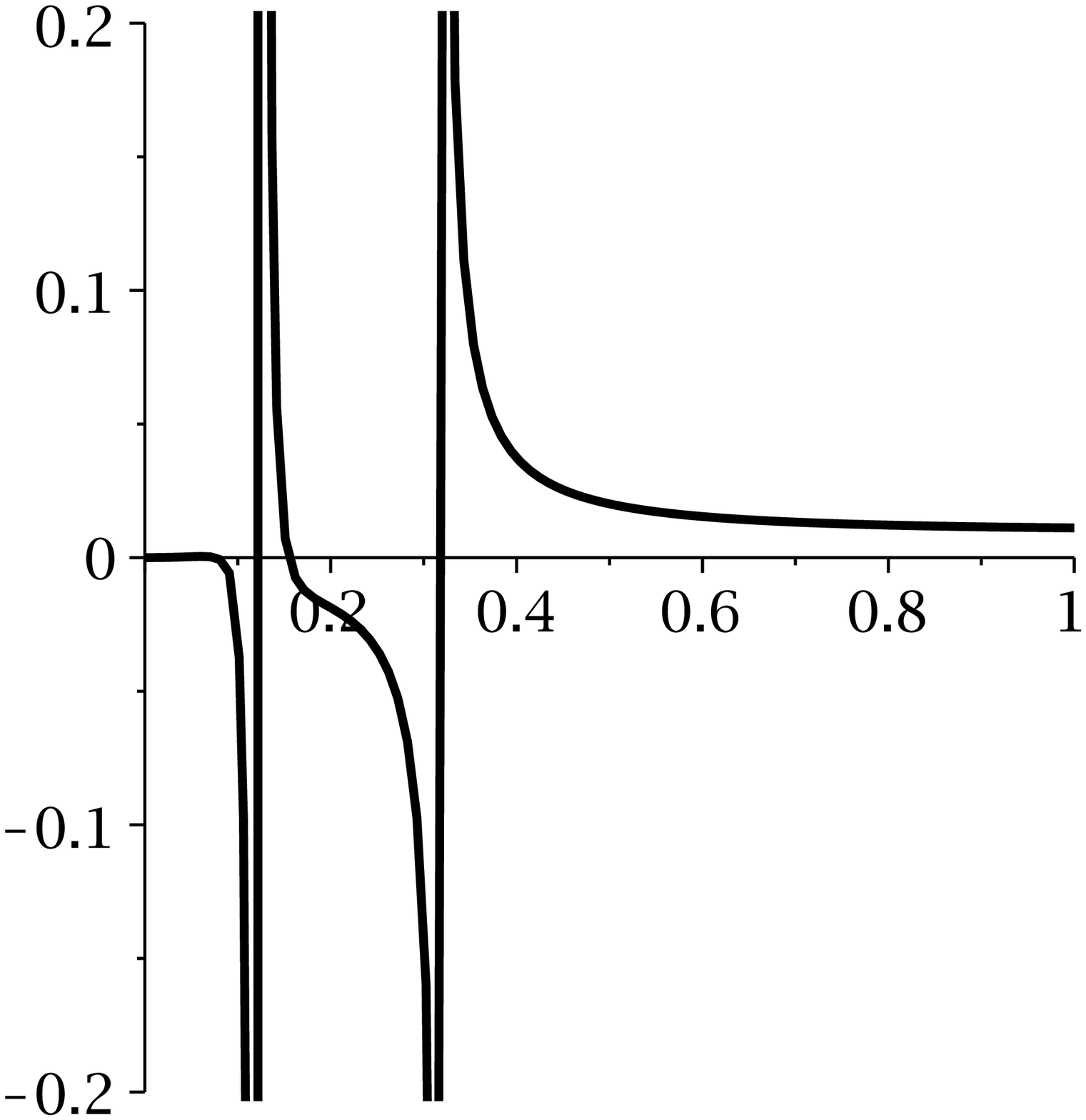} & \epsfxsize=5cm \epsffile{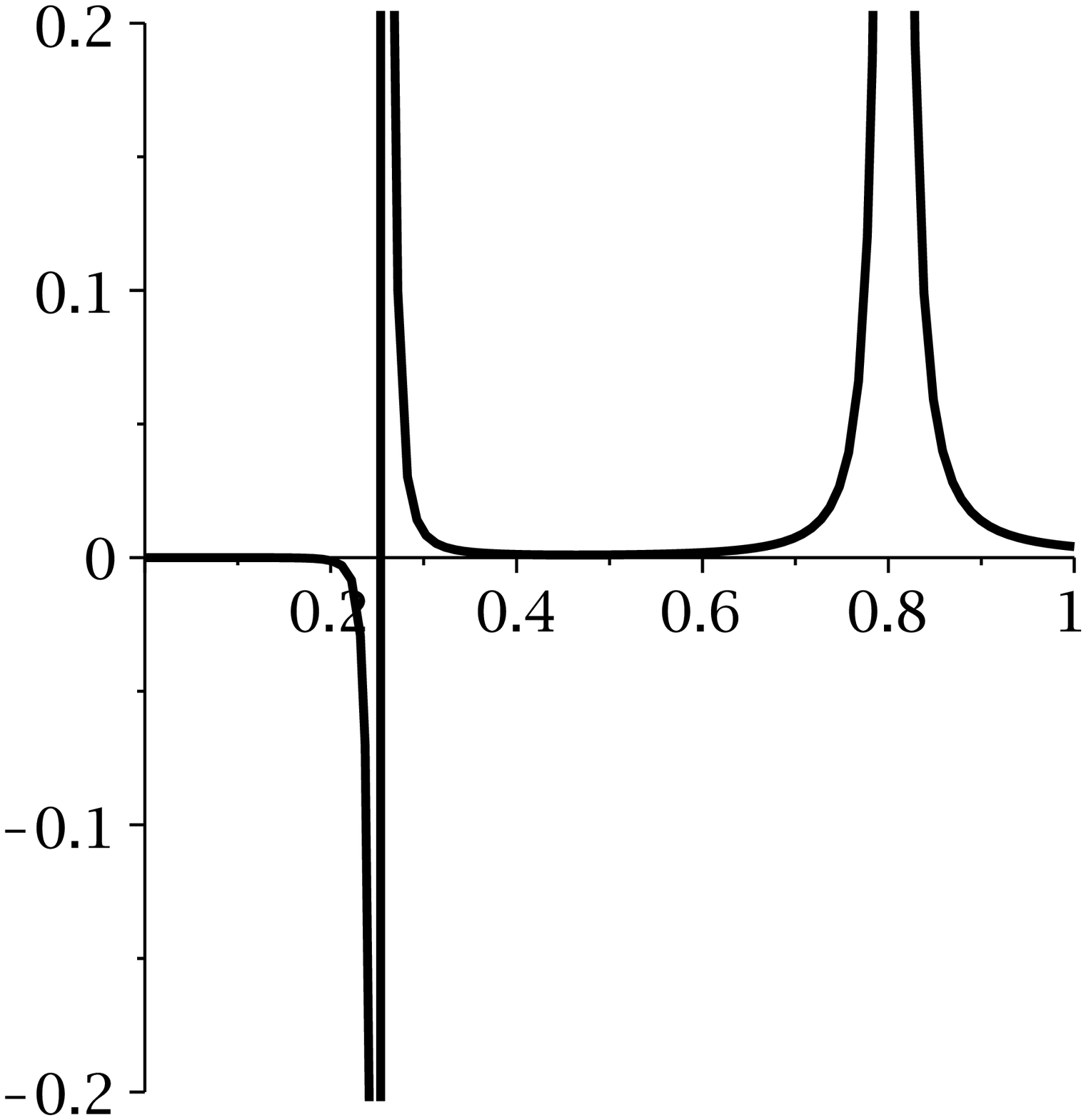}%
\end{array}
$%
\caption{ $\mathcal{R}$ versus $r_{+}$ for $q=2$, $b=1$, $\protect\alpha=5$,
$s=0.9$ and $\Lambda =-1$; \newline
left panel: Weinhold; middle panel: Ruppeiner; right panel: Quevedo.}
\label{Fig5}
\end{figure}

\begin{figure}[tbp]
$%
\begin{array}{cc}
\epsfxsize=5cm \epsffile{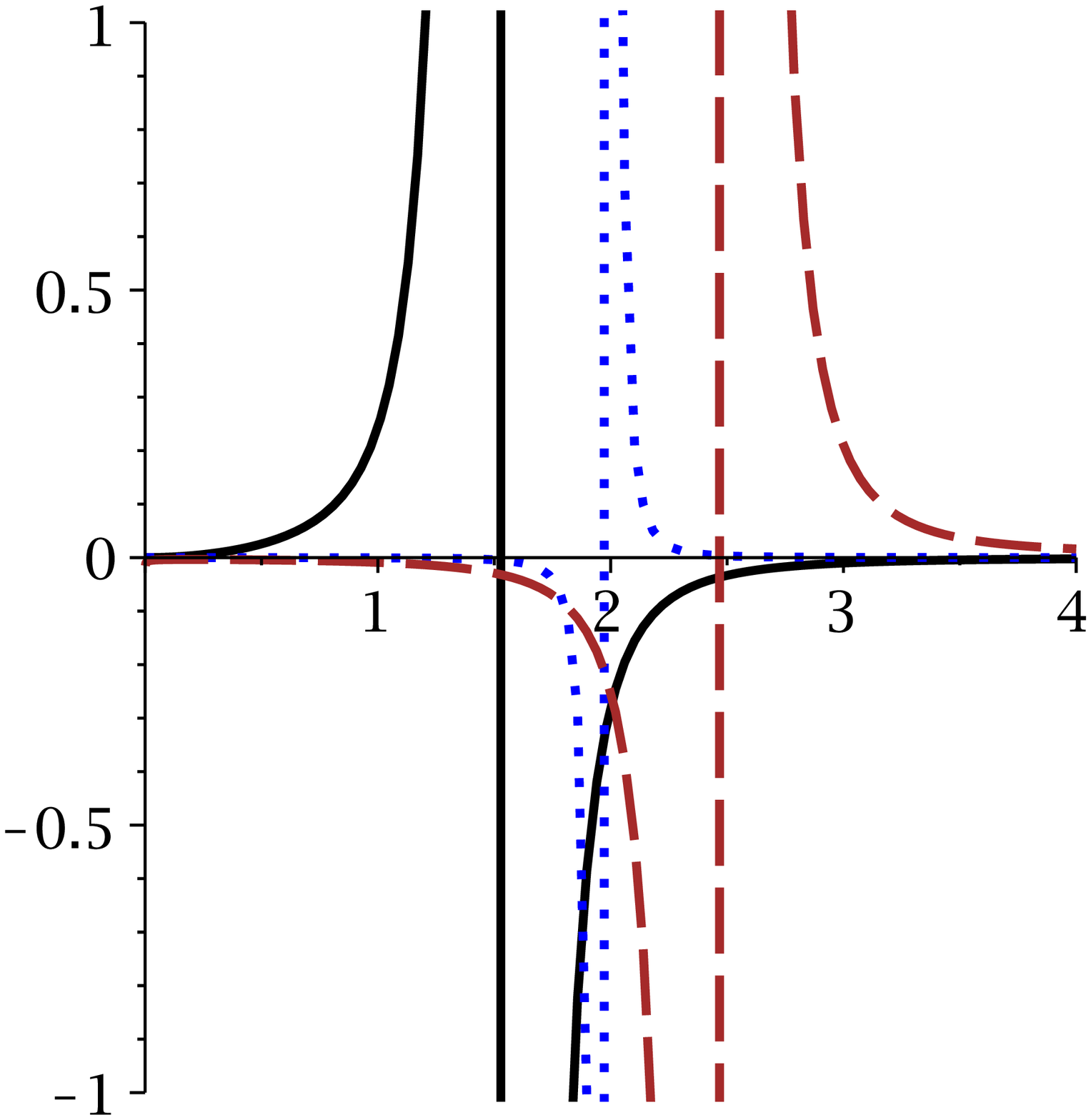} & \epsfxsize=5cm %
\epsffile{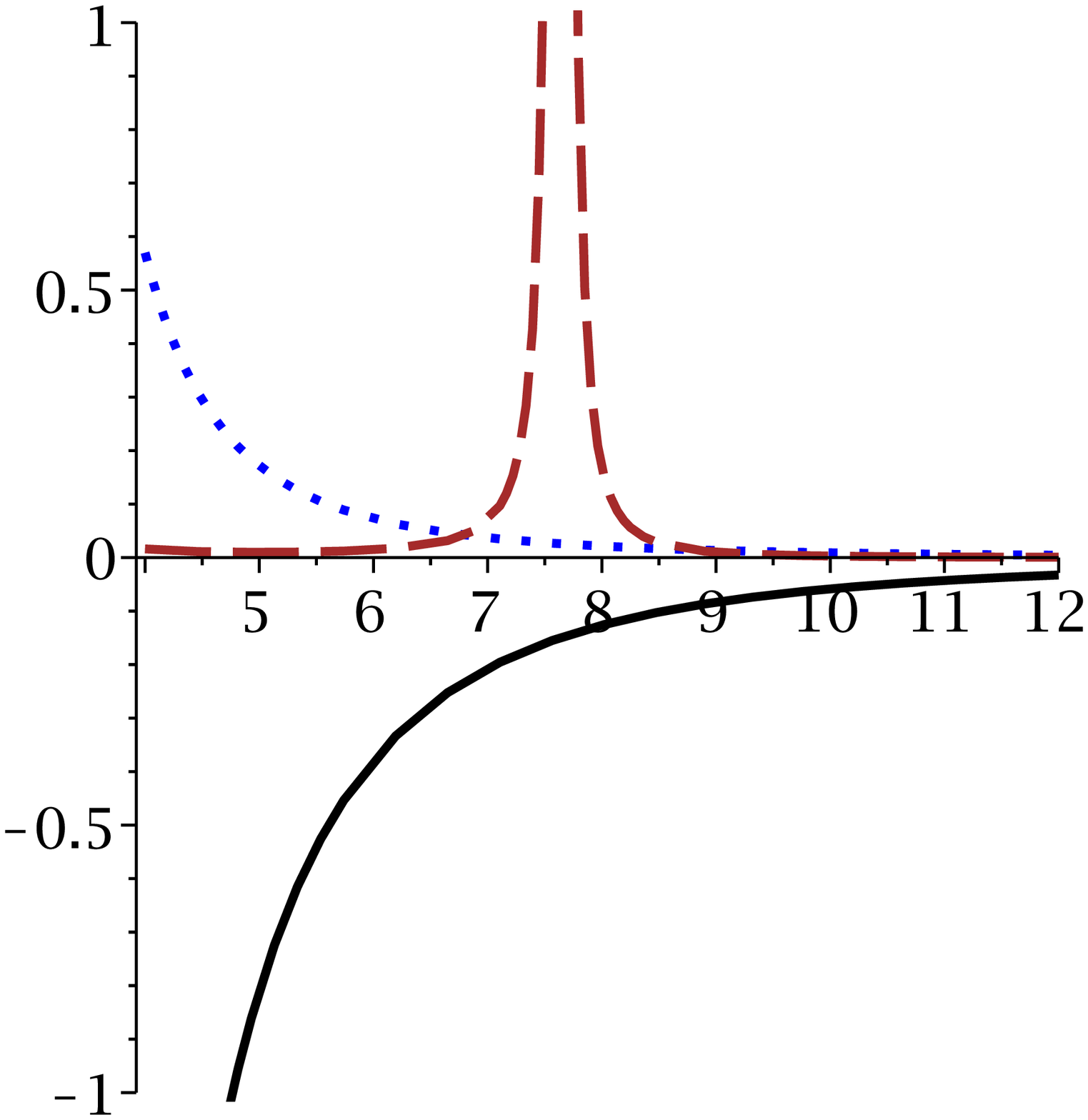} \\
\epsfxsize=5cm \epsffile{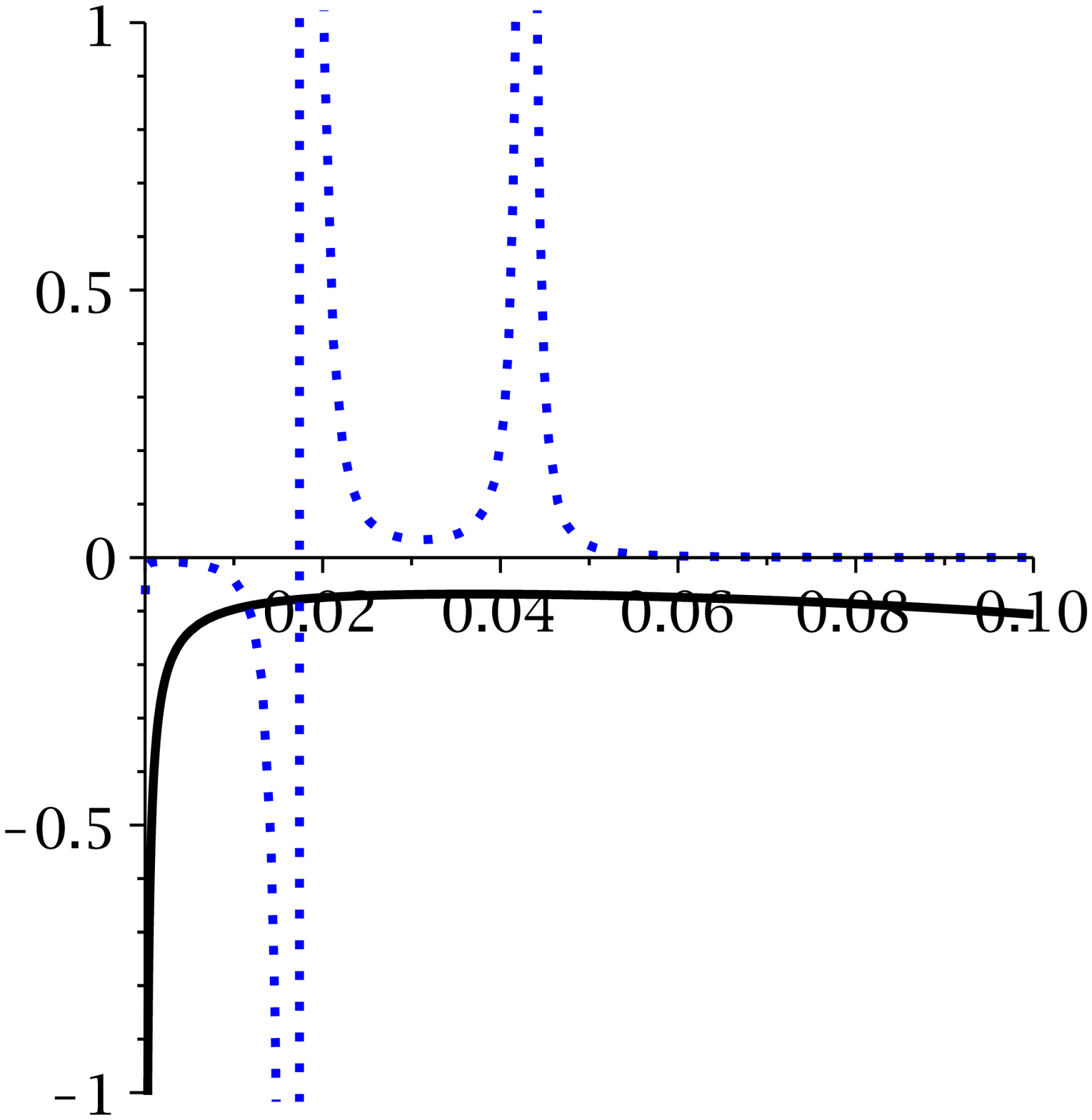} & \epsfxsize=5cm %
\epsffile{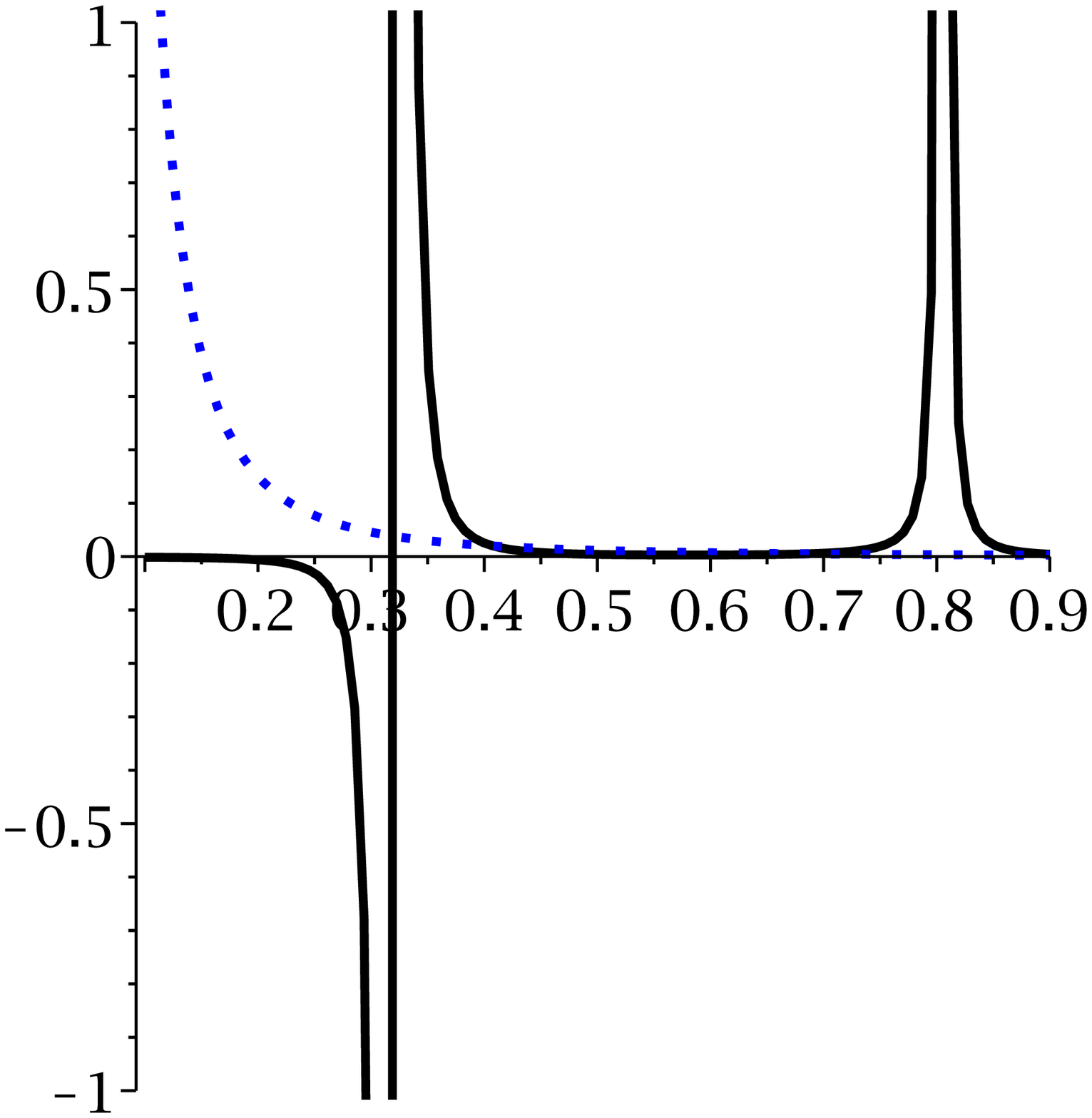}%
\end{array}
$%
\caption{For $\Lambda =-1$, $q=2$, $s=0.9$ and $b=1$. \newline
up panels: $\mathcal{R}$ versus $r_{+}$ diagrams for $\protect\alpha=0$
(continuous line), $\protect\alpha=1$ (dotted line) and $\protect\alpha=2$
(dashed line). \newline
down panels: $\protect\alpha=5$ (continuous line) and $\protect\alpha=8$
(dotted line). }
\label{Fig6}
\end{figure}

\begin{figure}[tbp]
$%
\begin{array}{ccc}
\epsfxsize=5cm \epsffile{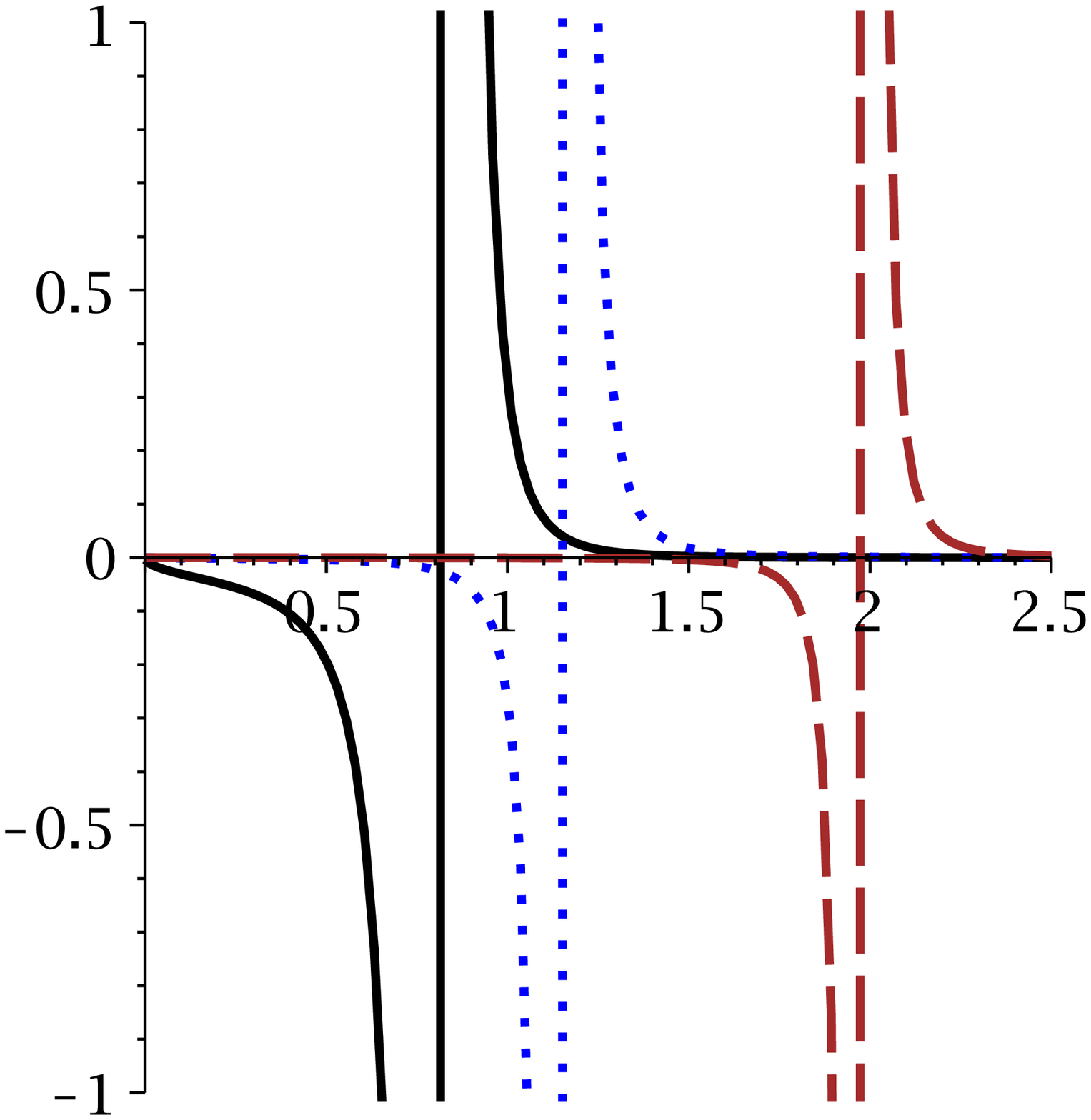} & \epsfxsize=5cm \epsffile{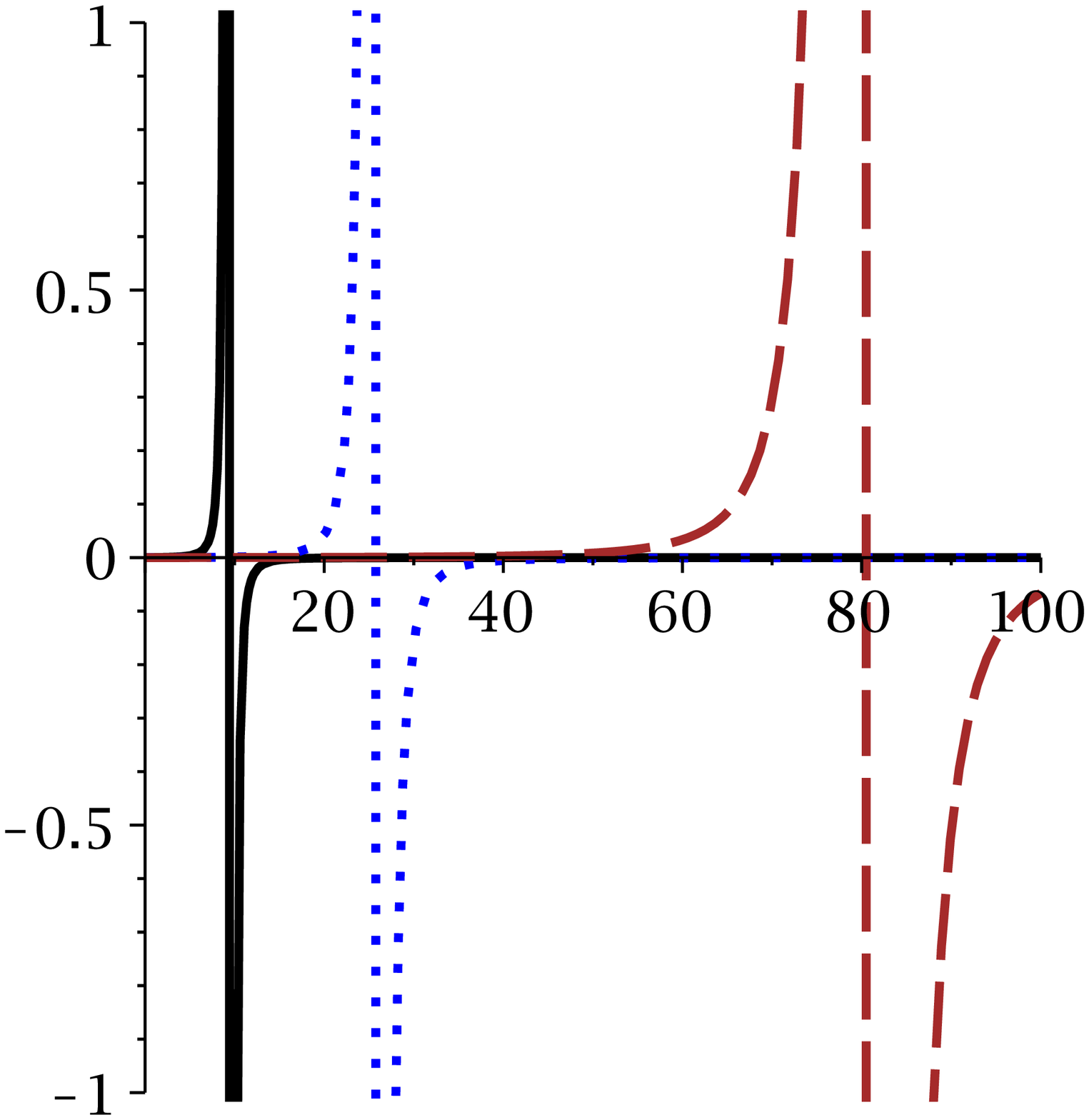}
& \epsfxsize=5cm \epsffile{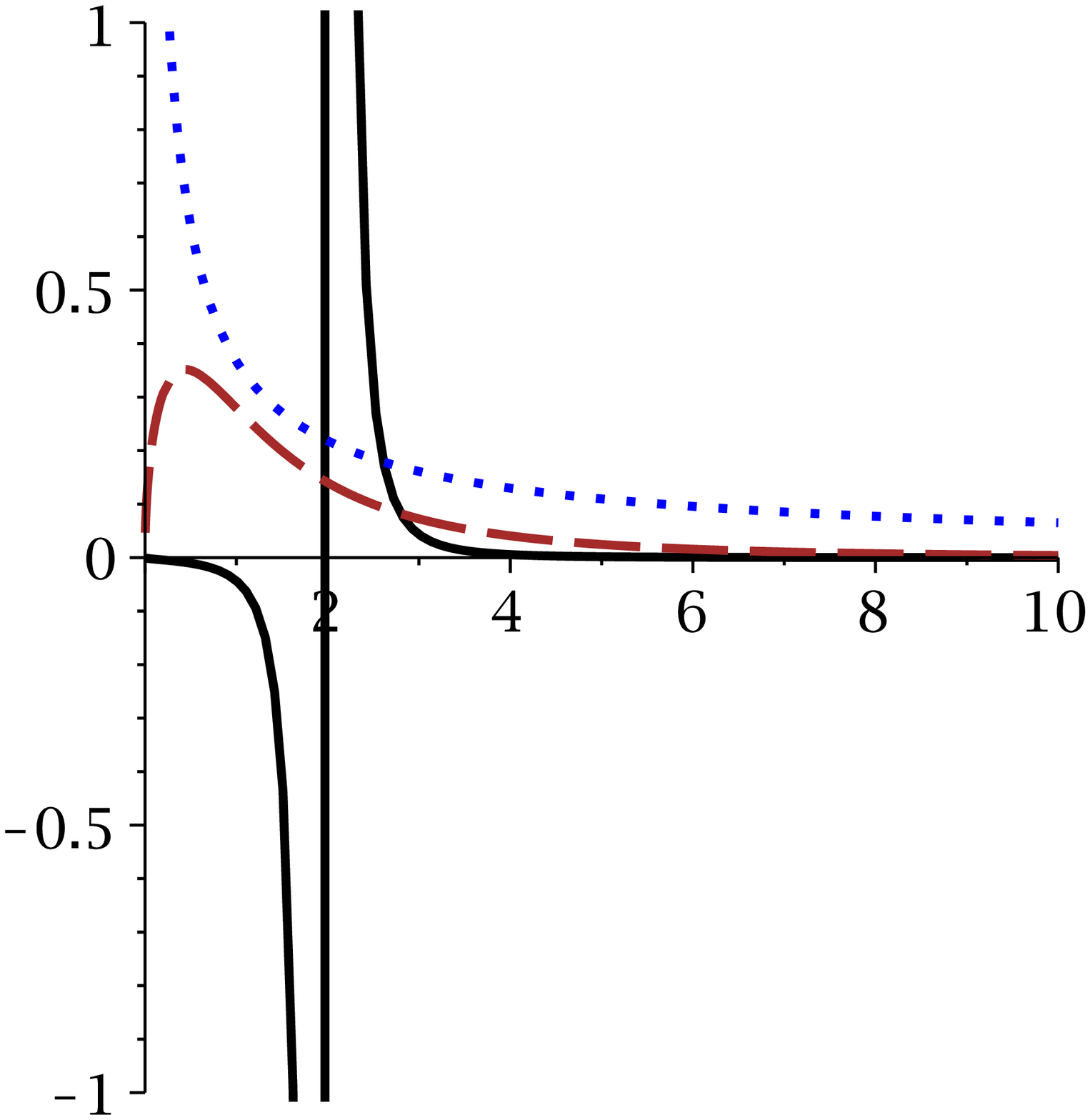}%
\end{array}
$%
\caption{ $\mathcal{R}$ versus $r_{+}$ for $q=2$, $\protect\alpha=1$ and $%
b=1 $; \newline
left panel: $\Lambda =-1$, $s=0.7$ (continuous line), $s=0.8$ (dotted line)
and $s=0.9$ (dashed line). \newline
middle panel: $\Lambda =-1$, $s=1.1$ (continuous line), $s=1.2$ (dotted
line) and $s=1.3$ (dashed line). \newline
right panel: $s=0.9$, $\Lambda =-1$ (continuous line), $\Lambda =0$ (dotted
line) and $\Lambda =1$ (dashed line).}
\label{Fig7}
\end{figure}


The mentioned thermodynamical metrics are in the following forms
\begin{equation}
ds^{2}=\left\{
\begin{array}{cc}
Mg_{ab}^{W}dX^{a}dX^{b} & Weinhold \\
&  \\
-T^{-1}Mg_{ab}^{R}dX^{a}dX^{b} & Ruppeiner \\
&  \\
\left( SM_{S}+QM_{Q}\right) \left( -M_{SS}dS^{2}+M_{QQ}dQ^{2}\right) &
Quevedo \\
&  \\
S\frac{M_{S}}{M_{QQ}^{3}}\left( -M_{SS}dS^{2}+M_{QQ}dQ^{2}\right) & HPEM%
\end{array}%
\right. ,
\end{equation}%
where $M_{X}=\partial M/\partial X$ and $M_{XX}=\partial ^{2}M/\partial
X^{2} $. It is a matter of calculation to show that the denominators of
Ricci scalar of these phase spaces are \cite{HPEMI}%
\begin{equation}
denom(\mathcal{R})=\left\{
\begin{array}{cc}
M^{2}\left( M_{SS}M_{QQ}-M_{SQ}^{2}\right) ^{2} & Weinhold \\
&  \\
M^{2}T\left( M_{SS}M_{QQ}-M_{SQ}^{2}\right) ^{2} & Ruppeiner \\
&  \\
M_{SS}^{2}M_{QQ}^{2}\left( SM_{S}+QM_{Q}\right) ^{3} & Quevedo \\
&  \\
2S^{3}M_{SS}^{2}M_{S}^{3} & HPEM%
\end{array}%
\right. .
\end{equation}

In order to have consistent results regarding bound and phase transition
points, the denominator of Ricci scalar of each metric must yield $M_{SS}$
and $M_{S}$, explicitly. Here, we see that HPEM metric has such factors in
its denominator of Ricci scalar. Whereas, the Quevedo metric has only $%
M_{SS} $ in its Ricci scalar's denominator and Weinhold and Ruppeiner do not
have these factors in explicit forms (see \cite{HPEMI} for more details). In
other words, the structures of denominators of the Ricci scalars obtained of
Weinhold, Ruppeiner and Quevedo metrics may yield extra divergencies for
their Ricci scalars which are not related to any bound and phase transition
points. In addition, these Ricci scalars may admit a mismatch between bound
and phase transition points and divergencies of the Ricci scalar. To
elaborate such cases, we have plotted Fig. \ref{Fig5}.

Evidently, for the Weinhold case , only one divergency for its Ricci scalar
is observed which is not matched with any bound and phase transition points
(compare left panel of Fig. \ref{Fig5} with down left panel of Fig. \ref%
{Fig1}). The Ruppeiner metric has two divergencies for its Ricci scalar; one
is matched with the bound point while the other one is not matched with any
phase transition point (compare middle panel of Fig. \ref{Fig5} with down
left panel of Fig. \ref{Fig1}). On the other hand, there are two
divergencies for Quevedo metric. One of them is coincidence with the phase
transition point while the other one is not matched with the bound point
(compare right panel of Fig. \ref{Fig5} with down left panel of Fig. \ref%
{Fig1}). It is evident that employing these three metrics leads into results
which are not consistent regarding the bound and phase transition points of
these black holes. In order to have consistent results, we employ the HPEM
metric and plot following diagrams (Figs. \ref{Fig6} and \ref{Fig7}).

A simple comparison between Figs. \ref{Fig6} and \ref{Fig7} with plotted
diagrams of the previous section (Figs. \ref{Fig1} and \ref{Fig4}) shows
that all the bound and phase transition points are matched with divergencies
of the Ricci scalar of HPEM metric for different cases. This confirms the
validation of the results of HPEM metric. Therefore, one can employ this
method as an independent approach for studying thermodynamical properties of
the black holes. This is the main purpose of the geometrical thermodynamics.
On the other hand, by taking a closer look, one can see that the sign of
HPEM Ricci scalar around the bound and phase transition points, depends on
the type of point. While we have a change of sign for Ricci scalar around
the bound point, but for the phase transition point, the sign of HPEM Ricci
scalar remains fixed. Therefore, it is possible to distinguish the type of
point by studying the sign of Ricci scalar of this metric. In the
geometrical thermodynamics, the sign of Ricci scalar determines whether the
system has repulsive or attractive interaction around the bound and phase
transition point. The positive sign indicates that system has repulsive
interaction whereas the opposite could be said about negative sign. Here, we
see that before the bound point, system has attractive property and by
crossing the bound point, the interaction is changed into repulsive. On the
other hand, for the phase transition point, the sign of Ricci scalar is
positive. Therefore, here, system has repulsive interaction on fundamental
level. Also, we see that employing the geometrical thermodynamics provides
us with extra information regarding the nature of interactions around the
bound and phase transition points. These information could not be extracted
by using the temperature and heat capacity of a system.


\section{Conclusion}

\label{Con}

We studied three dimensional dilatonic black holes in the presence of a
nonlinear electromagnetic field known as power Maxwell invariant. It was
shown that the solutions have interpretation of the black holes. Conserved
and thermodynamical quantities of these solutions were extracted and it was
shown that the first law of black holes thermodynamics is satisfied.

In order to highlight the importance and the role of mass (internal energy),
a thermodynamical investigation with the analyzing mass, temperature and
heat capacity of the black holes was done.

It was shown that it is possible to obtain roots for the mass of black holes
which led to the presence of region where the mass of black holes was
negative. This shows that in order to have a better picture regarding
thermodynamical structure of the black holes, it is necessary to include the
behavior of mass as well. There are specific restrictions which are imposed
by thermodynamical behavior of the mass which could not be neglected and
must be taken into account for having reliable conclusions and predictions
regarding thermodynamics of the black holes. For the past decades, the
studies regarding thermodynamical structures of the black holes have
neglected this important and crucial factor. Considering that we are using
thermodynamics of the black holes in AdS spacetime for applications in
conformal field theory, the restrictions which are imposed by mass of the
black holes become highly important.

In addition, the effects of different parameters on thermodynamical
structures and phase transitions of these black holes were investigated. It
was shown that the existence of phase transition is depending on the values
of different parameters.

Next, geometrical thermodynamics was employed to study thermodynamical
structure of these black holes. It was shown that for these specific black
holes, using Weinhold, Ruppeiner and Quevedo methods lead to inconsistent
results regarding the bound and phase transition points, whereas, HPEM
approach was successful in describing thermodynamical properties of these
black holes. Furthermore, we employed the sign of HPEM metric around the
bound and phase transition point to study repulsiveness and attractiveness
of the interactions.

Finally, it is worthwhile to generalize the obtained results to the case of
non-abelian Yang-Mills field. In addition, it will be interesting to study
causal structure of the solutions and examine the possibility of closed
timelike curves in the special case of dilaton and nonlinearity parameters.
Furthermore, one may pay attention to the possibility of having three
dimensional hairy black holes in the context of massive gravity. We will
address these subjects in the future works.

\begin{acknowledgements}
We thank the Shiraz University Research Council. This work has
been supported financially by the Research Institute for Astronomy
and Astrophysics of Maragha, Iran.
\end{acknowledgements}


\end{document}